\documentclass[11pt,a4paper]{article}
\usepackage[sort,numbers, square, comma, compress]{natbib}
\usepackage{amssymb}
\usepackage{amsthm}
\usepackage{amstext}
\usepackage{amsmath}
\usepackage{xfrac}
\usepackage{amsfonts}
\usepackage{empheq}
\usepackage{verbatim}
\usepackage{geometry}
\usepackage{latexsym}
\usepackage{feynmp-auto} 
\usepackage{caption}
\usepackage{subcaption}
\usepackage{color}
\usepackage{t1enc}
\usepackage{graphicx}
\usepackage[all]{xy}
\usepackage{makeidx}
\usepackage{slashed}
\usepackage{multicol}
\usepackage{alltt}
\usepackage{float}
\usepackage[hidelinks]{hyperref}

\allowdisplaybreaks

\newcommand{\be}{\begin{equation}}
\newcommand{\ee}{\end{equation}}

\newcommand{\bea}{\begin{eqnarray}}
\newcommand{\eea}{\end{eqnarray}}

\newcommand{\FDFI}{\left(\varphi^\dagger\overleftrightarrow{D}^I_\mu\varphi\right)}

\numberwithin{equation}{section}
\begin{document}
\begin{titlepage}
\hbox{NIKHEF/2018-031}
\hbox{CERN-TH-2018-157}

\vskip 30mm

\begin{center}
\Large{\sc{Effective operators in $t$-channel single top production\\[1ex] and decay}}
\end{center}

\vskip 6mm

\begin{center}

M.~de Beurs$^1$, E.~Laenen$^{1,2,3}$, M.~Vreeswijk$^1$, 
E. Vryonidou$^{1,4}$\\ [6mm]

\vspace{3mm}

\textit{$^1$Nikhef, Science Park 105, Amsterdam, The Netherlands} \\ 
\vspace{1mm}

\textit{$^2$ITFA, University of Amsterdam, Science Park 904, Amsterdam, 
The Netherlands} \\
\vspace{1mm}

\textit{$^3$ITF, Utrecht University, Leuvenlaan 4, Utrecht, The
  Netherlands} \\
\vspace{1mm}

\textit{$^4$CERN Theory Division, CH-1211 Geneva 23, Switzerland} \\
\vspace{1mm}

\end{center}

\vspace{0.5cm}

\begin{abstract}
  The production of a single top quark in the $t$-channel and its
  subsequent decay is studied at NLO accuracy in QCD, augmented with
  the relevant dimension-6 effective operators from the Standard Model
  Effective Theory. 
   We examine various kinematic and angular
  distributions for proton-proton collisions at the LHC at 13~TeV, in
  order to assess the sensitivity to these operators, both with and
  without the top quark narrow width approximation. 
  Our results will be helpful when devising strategies to establish bounds on their coefficients, including
 the amount of CP violation of the weak dipole operator.
\end{abstract}

\vskip -1mm

\noindent

\end{titlepage}
\tableofcontents

\section{Introduction}\label{sec:introduction}

Since its 1995 discovery \cite{Abe:1995hr,Abachi:1995iq} by the CDF
and D0 experiments the top quark has been an object of special
interest in high-energy physics. Its large mass $m_t$, the largest of
any known elementary particle, implies a strong coupling to the Higgs
mechanism, and ensures that QCD corrections, proportional to
$\alpha_s(m_t)$, are not overly large. The large mass also implies a
large width, mainly composed of decays to a $W$-boson and a bottom
quark, which prevents hadronisation, and enables clean transmission of
spin information to the decay products.  All of these characteristics
invite careful testing in diligent comparisons of experiment and
theory.  The study of the single top production process has the added
interest of directly involving the weak, charged-current interaction,
allowing stringent testing of its flavour-changing, chiral nature.

A key motivation behind such a precise scrutiny of the top quark is that
its production and decay mechanisms should be especially sensitive to
effects of physics beyond the Standard Model (SM).  A systematic
approach towards testing for the presence of such effects is the
framework of effective field theory, in which the Standard Model is
extended with higher-dimension operators that capture the effects of
new physics in a model-independent way
\cite{Zhang:2010dr,AguilarSaavedra:2008zc}:
\begin{equation}
  \label{eq:5}
  {\cal L}_{\mathrm{SM}} + \sum_i  \frac{C_i }{\Lambda^2} O^{[6]}_i +{\mathrm{hermitian\; conjugate}}
\end{equation}
where $\Lambda$ is the scale of new physics, typically taken to be
a few TeV, $O^{[6]}_i$ are
dimension-6 operators, and $C_i$ their associated coefficient
functions.  If one assumes that these operators
maintain SM symmetries one is lead to the SM Effective Field Theory
(SMEFT) \cite{Grzadkowski:2010es}.

One of the virtues of single top production in the $t$-channel (for massless $b$-quarks) is that at leading
order in QCD and at $\mathcal{O}(1/\Lambda^2)$ only three operators
$O^{[6]}_i$ with corresponding coefficients $C_i$ are required to
parameterise new physics effects: $O^{(3)}_{\varphi Q}$, $O_{tW}$ and $O^{(3)}_{qQ,rs}$:
\begin{eqnarray}
  O_{\varphi Q}^{(3)}
  &=&i \frac{1}{2} y_t^2 \FDFI (\bar{Q}\gamma^\mu\tau^I Q) \label{O1}
  \\
  O_{tW}&=&y_t g_w(\bar{Q}\sigma^{\mu\nu}\tau^It)\tilde{\varphi}W_{\mu\nu}^I
            \label{O2}  \\
  O^{(3)}_{qQ,rs}&=&\left(\bar q_r\gamma^\mu \tau^I q_s\right)\left(\bar Q\gamma_\mu \tau^I Q\right)  \label{O3}
\end{eqnarray}
where we have followed the notation and normalisation choice given in
\cite{Zhang:2016omx}, and dropped the superscript denoting that the operators are of dimension-6. These operators run and 
mix under renormalisation group evolution
\cite{Alonso:2013hga,Jenkins:2013wua,Jenkins:2013zja}, but we shall
omit these effects in our analysis.

Let us note here that more operators can
  contribute starting at $\mathcal{O}(1/\Lambda^4)$ such as the
  operators involving right handed bottom quarks, e.g. the dipole
  operator of the bottom quark, whose contributions are suppressed by
  the bottom mass at $\mathcal{O}(1/\Lambda^2)$. Four-fermion
  operators involving right-handed light quarks can also be relevant
  at $\mathcal{O}(1/\Lambda^4)$ \cite{Aguilar-Saavedra:2017nik}, but these are eliminated if one
  assumes Minimal Flavour Violation \cite{DAmbrosio:2002vsn}.
In general we assume the contribution from dimension-8 operators
to be sufficiently suppressed by their associated  $1/\Lambda^4$ prefactor. We shall
however use order $1/\Lambda^4$ contributions to the cross section
arising from squared contributions of dimension-6 ones to assess uncertainties.
Finally flavour changing
  interactions can also contribute to single top production, but we do
  not consider this here. For a recent global analysis of top-quark
  related flavour changing interactions in the effective operator
  framework see \cite{Durieux:2014xla}. 

This paper assesses the effect of the limited set of dimension-6
operators on single top quark production in the $t$-channel (for
brevity we show results for top production, but the same observations can
be made in anti-top production). We do so moreover at next-to-leading
order (NLO) in QCD, including top quark decay to $W$ and $b$, both in the narrow top 
width approximation (NWA), and by producing the $Wb$ directly, 
including non-resonant contributions. 

The paper is structured as follows. In the next section we discuss the necessary
background to single top production in SMEFT. In section 3 we present our
results, highlighting the opportunities in constraining the
dimension-6 operators with present and future LHC data. 
In the final section we present our conclusions.


\section{Single top production in SM extended to dimension-6}
\label{sec:single-top-prod}

\begin{figure}
  \centering \unitlength = 0.8mm
  \begin{fmffile}{t-channel1}
    \begin{fmfgraph*}(60,30) \fmfstraight \fmfleft{i1,i2}
      \fmfright{o1,o2} \fmf{fermion,tension=1,label=$u$}{i1,v1}
      \fmf{fermion,tension=1,label=$d$}{v1,o1}
      \fmf{fermion,tension=1,label=$b$}{i2,v2}
      \fmf{zigzag,tension=0,label=$W$}{v1,v2} 
      \fmf{fermion,tension=1,label=$t$}{v2,o2} \fmfv{d.sh=circle,
        d.fi=full, d.si=3thick, f=(1,,0,,0)}{v2}
    \end{fmfgraph*}
  \end{fmffile}
  \write18{mpost t-channel1} \unitlength = 0.8mm
  \begin{fmffile}{t-channel2}
    \begin{fmfgraph*}(60,30) \fmfstraight \fmfleft{i1,i2}
      \fmfright{o1,o2}
      \fmf{fermion,tension=1,label=$\bar{d}$,label.side=left}{v1,i1}
      \fmf{fermion,tension=1,label=$\bar{u}$,label.side=left}{o1,v1}
      \fmf{fermion,tension=1,label=$b$}{i2,v2}
      \fmf{zigzag,tension=0,label=$W$}{v1,v2} 
      \fmf{fermion,tension=1,label=$t$}{v2,o2} \fmfv{d.sh=circle,
        d.fi=full, d.si=3thick, f=(1,,0,,0)}{v2}
    \end{fmfgraph*}
  \end{fmffile}
  \write18{mpost t-channel2} \vspace{0.5cm}
  \caption{Feynman diagrams for $t$-channel single top production at
    LO.  An incoming bottom quark and either an incoming up type
    (left) or anti-down type (right) quark exchange a virtual $W$
    boson, a so-called $t$-channel exchange. The outgoing $d$ or
    $\bar{u}$ quark can be observed as a jet. The red vertex
    corresponds to eq. \ref{eq:1}, and allows, according to the SM,
    only left-handed top quarks to be produced.}
  \label{tchannel}
\end{figure}
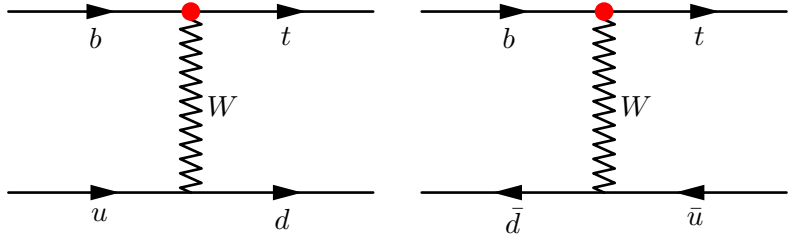

To establish our context we recall here some basic aspects of single top
production, and the associated charged current interaction. 
The leading order diagrams for $t$-channel single top production are shown
in fig.~\ref{tchannel}.

The essence of the single top production channel in the SM is that the top quark
is produced through an interaction with a $W$ boson. This interaction
corresponds to the following term in the SM Lagrangian
\begin{equation}
  \label{eq:1}
  {\cal L}^{\mathrm{SM}}_{Wtb} = -\sum_{f=d,s,b}^3\frac{gV_{tf}}{\sqrt{2}}\,  \bar{q}_f(x) \gamma^\mu 
  P_L t(x)\,W_{\mu}(x)  + {\mathrm{h.\;c.}}\,,
\end{equation}
The coupling strength is denoted by $g$, and top quark $t(x)$ and
$W$-boson $W_\mu(x)$ fields are indicated, as are quark fields
$q_f(x)$, where $f=d,s,b$ indicates down, strange or bottom
quarks. The coefficient $V_{tf}$ is an element of the
Cabibbo-Kobayashi-Maskawa (CKM) matrix.
Also shown is the projection operator $P_L$ which projects onto the
left-handed (V-A) part of the top quark.  Once produced, the top (or
anti-top) quark decays almost always into a $b$ quark and a $W$ boson
which subsequently decays to a positron (or electron) and the
corresponding (anti-)neutrino.\footnote{ Only electronic $W$ decays are used in this study. Similar results are expected for top
  quark events with a muon in the final state. Events where the $W$ is decaying to a $\tau$, or hadronically, are experimentally more difficult to isolate. }  The notable aspect of this decay is that
there is a near-perfect correlation between the flight direction of
the positron in the top quark rest frame, and the top quark spin
\cite{Mahlon:1996pn,Mahlon:1999gz}.  As the positron is easily detected, this correlation
allows a direct determination of the top quark spin, and the
handedness of the coupling from the positron angular distribution.

Any new physics altering the $Wtb$ interaction can then be probed by
studying single top production and decay. The SMEFT can parameterise
deviations from the SM predictions and can be used to make
quantitative predictions to be compared with experimental data.

The operators of eq.~\eqref{O1}-\eqref{O2} modify the $Wtb$
interaction in the following way
\begin{eqnarray}
  {\cal L}^{\mathrm{dim-6}}_{Wtb} &=& -\frac{g}{\sqrt{2}}  \bar{b}(x) \gamma^\mu 
                                      P_L t(x)\,W_{\mu}(x) \left(1+  \frac{C^{(3)}_{\varphi Q} y_t^2 v^2}{2\Lambda^2}  \right) \nonumber \\
                                  &+&  \frac{2 \,g \,v \,y_t \, C_{tW}}{\Lambda^2}  \bar{b}(x) \sigma^{\mu\nu} 
                                      P_R t(x)\,\partial_{\nu} W_{\mu}(x) +  {\mathrm{h.\; c.}} \,,\label{interaction}
\end{eqnarray} 
where $v = 246$ GeV is the Higgs doublet vacuum expectation value, and
$y_t$ the top quark Yukawa coupling. Here and below we assume
$V_{tb}=1$.  Note that the four-fermion operator of eq.~\eqref{O3} introduces a
contact $udtb$ interaction.  The impact of these operators can be
already seen by considering the partonic single top cross section,
which at $\mathcal{O}(1/\Lambda^2)$ can be written schematically as
\begin{equation}
  \label{eq:3}
  \frac{d\sigma_{ub\rightarrow dt}}{d \cos\theta} = \left(1 + \frac{C^{(3)}_{\varphi Q } y_t^2  v^2}{\Lambda^2} \right)  k_1( \theta) +
  \frac{C^{(3)}_{qQ,rs}}{\Lambda^2}  k_2(\theta) +
  \frac{\mathrm{Re\,}C_{tW}}{\Lambda^2}  k_3(\theta)\,,
\end{equation}
where the $k_i$ are known functions of $\theta$, the angle
between the incoming bottom quark direction and the top quark flight
direction in the partonic center-of-mass frame.

An interesting feature of this production cross section is that each
of the coefficients $C^{(3)}_{\varphi Q}$, $C^{(3)}_{qQ,rs}$ and
$\mathrm{Re\,} C_{tW}$ is associated with a specific angular
dependence, enabling one to determine, or at least bound, the
individual contributions experimentally.
\begin{figure}
  \begin{center}
    \includegraphics[scale=0.5,page=1]{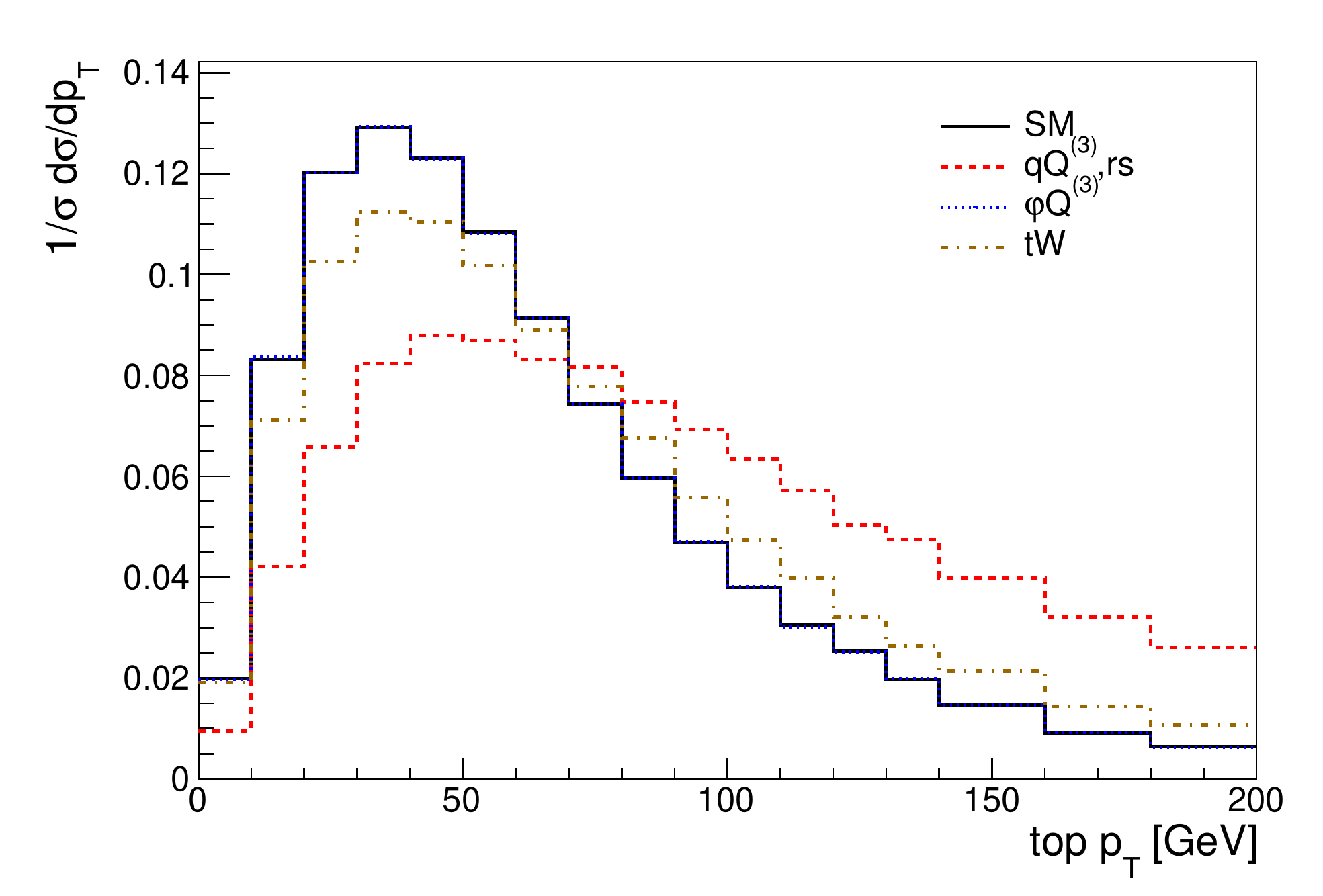}
  \end{center}
  \vspace{-0.5cm}
  \caption{ The normalised leading order parton-level differential cross
    section as a function of the transverse momentum of the top quark.
     The expectation of the SM together with the
    interference effects of the three effective operators of interest are
    shown.}
  \label{production}
\end{figure}
The operator $O^{(3)}_{\varphi Q}$ only modifies the magnitude of the
$Wtb$ interaction as shown in eq.~\eqref{interaction}, but does not
change the angular dependence of the SM prediction.  By contrast, the
operator $O^{(3)}_{qQ,rs}$, with corresponding real coefficient
$C^{(3)}_{qQ,rs}$, represents a four-quark contact interaction and
noticeably affects the angular distribution of the top quark
production angle. Of course, eq.~\eqref{interaction} addresses only
the dominant, lowest order parton process $u+b\rightarrow d+t$.  Other
partonic processes also contribute but the different angular behaviour of
the partonic cross section predicted by the different operators
directly translates into different shapes of the top transverse
momentum distribution. This is illustrated in fig.~\ref{production},
where the effect of $C^{(3)}_{qQ,rs}$ on the top $p_T$ distribution is
clearly distinguishable. Finally, the contribution of
$\mathrm{Re\,}C_{tW}$ has a signature again different from the other
two operators, but its effect is smaller and is better determined in
the decay of top quarks than in their production \cite{Zhang:2016omx}.

The above discussion is somewhat simplified as it refers to the lowest
order contributions in both QCD and the EFT expansion. Next-to-leading
order (NLO) QCD corrections can be also relevant and can potentially
modify the relative contributions of the operators. At NLO in QCD, the
chromomagnetic dipole operator, $O_{tG}$, contributes as discussed in
\cite{Zhang:2016omx} whilst additional operators contribute at
$\mathcal{O}(1/\Lambda^4)$.  We omit these operator contributions in
this work but in future work we intend to take them into
consideration.  \footnote{We remark that a slightly different approach
  \cite{AguilarSaavedra:2006fy,AguilarSaavedra:2008gt,Aguilar-Saavedra:2014eqa},
  not using operators but anomalous couplings, has also been used in
  the literature. The connection between the operator coefficients to
  the anomalous couplings is discussed in
  \cite{AguilarSaavedra:2008zc}. Here all types of Lorentz-invariant
  interaction structures that involve the $W$ boson and the top quark
  are allowed, including those that the Standard Model does not
  allow. An advantage of the present approach is the limited number of
  parameters, a restriction following from symmetry requirements.}

Given the different angular distributions already observed at the
level of the partonic cross section, it is interesting to fully
explore $t$-channel single top production in the presence of the
dimension-6 operators and provide the relevant predictions for the
LHC. In the following sections we will therefore study single top
production in the presence of the dimension-6 operators in eq.~(\ref{O1} -
\ref{O3}) both at the inclusive and differential level, as well as including top decays,
and we will identify observables that can be used to bound the values of
the corresponding coefficients $C^{(3)}_{\varphi Q}$,
$C^{(3)}_{qQ,rs}$ and $C_{tW}$.


\section{Numerical studies}
\label{sec:numerical-studies}
To study the impact of the three operators on single
top production we compute the corresponding contributions at LO and NLO 
matched to the parton shower (PS). The computation is performed within the
{\sc MadGraph5\_aMC@NLO} ({\sc MG5\_aMC}) framework
\cite{Alwall:2014hca}, and uses the NLO EFT implementation of
Ref.~\cite{Zhang:2016omx}.  While \cite{Zhang:2016omx} produces
results for stable top quarks, we will also consider the top quark
decays. This can by achieved by either decaying the top-quark in {\sc MadSpin} \cite{Artoisenet:2012st} 
or by following the procedure of resonance-aware PS matching
presented in \cite{Frederix:2016rdc}, to produce a $Wbj$ final state.
By decaying the $W$ boson in {\sc MadSpin}, we retain
spin information.
Our setup is fully differential and allows us to assess the impact of
NLO corrections as well as the impact of the operators entering
either in the production or in the decay, or both, for any
observable. 

To facilitate discussion we first fix our notation.
Assuming one insertion of each operator,
we can write the matrix element for single top production in the form
\begin{equation}
  \mathcal{M} = \mathcal{M}_{\text{SM}} + \sum_i\frac{1{\rm TeV}^2}{\Lambda^2}C_i \,\mathcal{M}_i\,,
\end{equation}
where the $\mathcal{M}_i$ are defined as having precisely one
insertion of operator $O_i$ in all possible ways. We have normalised
the new physics scale $\Lambda$ in units of TeV.  In physical
observables, such as the production cross section and the top width, the
matrix element enters squared. The squared amplitude takes the form:
\begin{equation}
  |\mathcal{M}|^2 = \mathcal{|M|}_{\text{SM}}^2 + \sum_i\frac{1{\rm
      TeV}^2}{\Lambda^2}C_i \, 
2\mathrm{Re}\left(\mathcal{M}^{*}_{\text{SM}} \, \mathcal{M}_i \right) 
+ \sum_{i\leq j} \frac{1{\rm TeV}^4}{\Lambda^4}C_iC_j \, \mathcal{|M|}^2_{i,j} \,,
  \label{eq:m_squared}
\end{equation}
assuming for simplicity real operator coefficients.  From here onwards the contribution
to the cross section from the
interference term with the SM
($\propto 2\mathrm{Re}\left(\mathcal{M}^{*}_{\text{SM}} \,
  \mathcal{M}_i \right)$) will be denoted by $\sigma_i$, while the
additional squared terms ($\propto\mathcal{|M|}^2_{i,j}$) will be
denoted by
$\sigma_{i,j}$. In this notation, the cross section can be parameterised as:
\begin{flalign}
	\sigma=\sigma_{\text{SM}}+\sum_i\frac{1{\rm TeV}^2}{\Lambda^2}C_i \, \sigma_i
	+\sum_{i\leq j}
	\frac{1{\rm TeV}^4}{\Lambda^4}C_iC_j \, \sigma_{i,j}\,.
	\label{eq:xsecpara}
\end{flalign}
We will present results for all three terms. 
We recall here our remark in section \ref{sec:introduction} that
the ${\cal O}(1/\Lambda^4)$ terms represented by the $\sigma_{i,j}$ are far
from complete, and we use them here only to estimate uncertainties in the EFT expansion.


\subsection{Inclusive single top production}
\label{sec:tj}
We start by computing the total single top production cross section
for stable top quarks for the relevant operators at LO and NLO for the
LHC, at 13 TeV. These results are also available in
Ref.~\cite{Zhang:2016omx}, but we reproduce them here in table
\ref{tab:tj1} for completeness. Our computation uses the
five-flavour number scheme.  For these results the renormalisation and
factorisation scales, $\mu_{R}$ and $\mu_F$ are both set to
$m_t = 172.5$ GeV.  The NNPDF3.0 LO and NLO sets \cite{Ball:2014uwa}
are used for the LO and NLO predictions respectively and the only
kinematic cuts are applied to the jets: $p_T^j > 5$ GeV and $|\eta_j|$ < 5. To show 
the impact of the NLO
corrections, table \ref{tab:tj1} presents the
$K$-factors which are defined as the ratio
$\sigma_{\textrm{NLO}}/\sigma_{\textrm{LO}}$ for each contribution.\\
\begin{table}
\renewcommand{\arraystretch}{1.8}
 \makebox[\linewidth]{
	\begin{tabular}{l|ll|ll|l}
& \multicolumn{2}{|c|}{LO} & \multicolumn{2}{|c|}{NLO} &  \\
Operator & $\sigma$ [pb] & $\frac{\sigma}{\sigma_{\text{SM}}}$ [\%] & $\sigma$ [pb] & $\frac{\sigma}{\sigma_{\text{SM}}}$ [\%] & $K$ \\
\hline
$\sigma_{\text{SM}}$ & 123 & - & 137 &  - & 1.12\\
$\sigma_{qQ,rs^{(3)}}$ & -92.3 & -75.3 & -102 & -74.7 & 1.11\\
$\sigma_{\varphi Q^{(3)}}$ & 14.6 & 11.9 & 16.3 & 11.9 & 1.12\\
$\sigma_{tW}$ & 3.05 & 2.49 & 3.57 & 2.6 & 1.17\\
$\sigma_{itW}$ & - & - & - & - & -\\
$\sigma_{qQ,rs^{(3)},\,qQ,rs^{(3)}}$ & 77.3 & 63.1 & 80.8 & 58.9 & 1.05\\
$\sigma_{\varphi Q^ {(3)},\varphi Q^ {(3)}}$ & 0.434 & 0.354 & 0.485 & 0.354 & 1.12\\
$\sigma_{tW,tW}$ & 0.758 & 0.619 & 1.03 & 0.752 & 1.36\\
$\sigma_{itW,itW}$ & 0.761 & 0.616 & 1.03 & 0.752 & 1.35\\
$\sigma_{qQ,rs^{(3)},\,\varphi Q^ {(3)}}$ & -5.49 & -4.48 & -6.08 & -4.43 & 1.11\\
$\sigma_{qQ,rs^{(3)},\,tW}$ & -2.34 & -1.91 & -2.84 & -2.07 & 1.22\\
$\sigma_{\varphi Q^ {(3)},tW}$ & 0.182 & 0.148 & 0.212 & 0.155 & 1.17\\	\end{tabular}}
\caption{\label{tab:tj1}
  Contributions to the cross section in pb for $t$-channel single top
  production at 13 TeV, as parameterised in eq.~(\ref{eq:xsecpara}). 
These values have been extracted from fitting eq.~(\ref{eq:xsecpara}), 
to a hundred computed cross sections with randomly chosen coupling
strengths for the effective operators, both for LO and NLO
separately. The statistical errors for each contribution in the table
is below 1\% except for the $\sigma_{qQij,tW}$ term at NLO, which is
at 1.1\%. The right-hand-side column shows the $K$-factor, which is 
defined for each row  as the ratio of the NLO over the LO
prediction. By the subscripts $tW$ and $itW$ we denote the contributions of the real and 
imaginary parts of $C_{tW}$ respectively.}
\end{table}
We find that for the single top process the squared terms and
interference between the operators, i.e. the
$\mathcal{O}(1/\Lambda^4)$ terms, are suppressed for coefficients of
$\mathcal{O}(1)$ for the $O_{tW}$ and $O_{\varphi Q^{3}}$ operators
but are not negligible for the 4-fermion operator. Taking its
coefficient to be of order one we find a large cancellation between the
interference and squared contributions. We also observe that
$K$-factors vary considerably between the various operators, and 
can be quite different from the SM contribution. 
This underlines the importance of including genuine NLO corrections in predictions,
since a universal $K$-factor does not summarise the table. In the table 
we also include the contribution of the imaginary part of the coefficient 
$C_{tW}$, which only enters squared at $\mathcal{O}(1/\Lambda^4)$ as it does not 
interfere with the SM or the other operators. We will discuss this contribution in detail in 
section \ref{sec:cp-odd}.
\\

\begin{table}
\renewcommand{\arraystretch}{1.8}
 \makebox[\linewidth]{
	\begin{tabular}{l|c|cc|cc|}
& & \multicolumn{2}{|c|}{LO} & \multicolumn{2}{|c|}{NLO}\\
Operator & Coupling value & $\sigma$[pb] $\pm$scale $\pm$PDF & $\Gamma_{\text{top}}$ [GeV] & $\sigma$[pb] $\pm$scale $\pm$PDF & $\Gamma_{\text{top}}$ [GeV] \\
\hline
$\text{SM}$ &  - & $123^{+9.3\text{\%}}_{-11.4\text{\%}} \pm 8.9\text{\%}$ & 1.49 & $137^{+2.7\text{\%}}_{-2.6\text{\%}} \pm{1.2\text{\%}}$ & 1.36\\
$O^{(3)}_{qQ,rs}$ & -0.4 & $172^{+8.7\text{\%}}_{-10.8\text{\%}} \pm 8.9\text{\%}$ & 1.49 & $190^{+2.4\text{\%}}_{-1.8\text{\%}} \pm 1.1\text{\%}$ & 1.35 \\
$O_{\varphi Q}^{(3)}$ & 1 & $137^{+9.3\text{\%}}_{-11.4\text{\%}} \pm 8.9\text{\%}$ & 1.67 & $154^{+2.3\text{\%}}_{-2.3\text{\%}} \pm 1.2\text{\%}$ & 1.52\\
$O_{tW}$ (Re) & 2 & $132^{+9.3\text{\%}}_{-11.4\text{\%}} \pm 8.8\text{\%}$ & 1.83 & $148^{+2.3\text{\%}}_{-2.5\text{\%}} \pm 1.2\text{\%}$ & 1.68\\
$O_{tW}$ (Im) & 1.75i & $125^{+9.2\text{\%}}_{-11.4\text{\%}} \pm 8.8\text{\%}$ & 1.51 & $140^{+2.3\text{\%}}_{-2.5\text{\%}} \pm 1.2\text{\%}$ & 1.38\\
\end{tabular}}
\caption{The benchmark choices for the coupling values of the effective operators, together with the corresponding $t$-channel single top cross section and the width of the top quark. The scale and PDF uncertainties of the cross sections are also shown.}
\label{tab:cvalues}
\end{table}



Total cross-section results give a good first indication on the impact of the operators
 on the single top production process, but more information can be extracted by
  considering differential distributions. To demonstrate the effect of the operators 
  on the differential distributions we select a set of benchmark scenarios. 
  The benchmark coupling values that will be used throughout the paper 
  are presented in table \ref{tab:cvalues}. We follow the EFT analyses of 
  Refs.~\cite{Zhang:2016omx,Cirigliano:2016nyn} to ensure that our coupling 
  values fall within the current limits. The effects on the inclusive cross section and the 
  top width are also given for both LO and NLO. The predicted deviations 
  from the SM predictions lie within the uncertainty of recent 
  single top measurements: $\sigma = 156 \pm 35$ pb and $0.6 \le \Gamma_{\text{top}} \le 2.5$ GeV \cite{Aaboud:2016ymp,Sirunyan:2016cdg,Aaboud:2017uqq,CMS-PAS-TOP-16-019}. 
  In the table we also include the scale uncertainties obtained by varying the central 
  renormalisation and factorisation scale by a factor of two up and down, and the PDF uncertainties. 
  We note the significant decrease in the scale 
  and PDF uncertainties going from LO to NLO, a well-known feature of NLO computations.
   At NLO the combined uncertainty is  only of the order of 3\%, in
   agreement with previous results \cite{Alwall:2014hca}. We shall therefore refrain from showing uncertainty bands
   in our differential distributions, even though these can be straightforwardly computed with our setup.

\begin{figure}[H]
\centering
\begin{subfigure}{0.5\textwidth}
\centering
\includegraphics[scale=0.4,page=1]{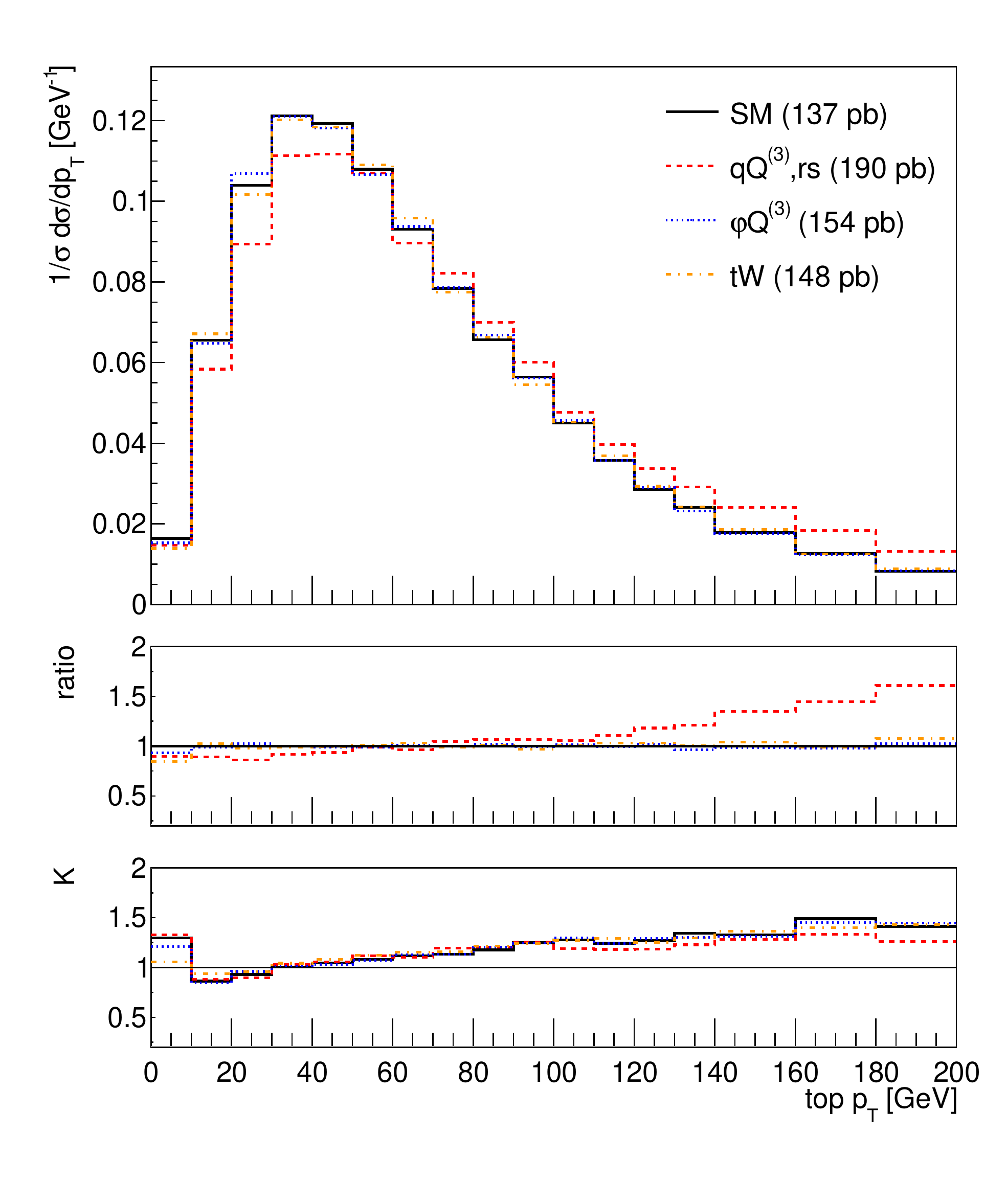}
\end{subfigure}%
\begin{subfigure}{0.5\textwidth}
\centering
\includegraphics[scale=0.4,page=2]{tj_plots.pdf}
\end{subfigure}
\caption{
The NLO distributions of the stable top quark transverse momentum and rapidity for the SM and the three
  effective operators of interest Eqs. (\ref{O1}-\ref{O3}) for the couplings values of table \ref{tab:cvalues}. 
  The ratio shown in the first inset is defined as the
  effect of the operator over the SM, the second inset shows the
  $K$-factor, the ratio of the NLO over the LO predictions.}
\label{fig:production_pt+eta}
\end{figure}

We start by showing the stable\footnote{\label{note:stable}This top is
  selected based on its particle ID (i.e. in this example it is not
  reconstructed from its decay products), and therefore stable.} top
quark transverse momentum and pseudorapidity distributions for SM and the first
three benchmarks of table \ref{tab:cvalues} in
fig.~\ref{fig:production_pt+eta}. Computing these distributions, we
allow only one operator coefficient to be non-zero at a time. We include the interference with the SM as well as the square
 terms. In the distributions we do not include the benchmark with imaginary $C_{tW}$ coefficient which will be discussed 
in detail in section \ref{sec:cp-odd}. 

In the distributions we see that the 4-fermion
operator in particular has an effect on the shapes in both the transverse momentum and
rapidity distributions, leading to harder and more central tops. The
impact of the other two operators on the distribution shapes is
milder. It can also be observed that the shape difference between LO
and NLO has its largest effect in the sensitive region of these
distributions, highlighting again the importance of NLO predictions for
experimental analyses of this process.


\subsection{Single top production and decay}
To study the process in more detail and extract maximal information on
the impact of the operators, we should consider the
distributions of the top decay products. This requires
studying the full process of $p p \to b \ell \nu j$, where we have
assumed that the top quark decays leptonically. In such a computation several difficulties
arise compared to that for the inclusive $pp \to tj$ computation. 

The first is that to generate a consistent single top event sample, the full
process $p p \to b \ell \nu j$ has to be generated, including 
both the off-shell top effects and the interference with all the irreducible
backgrounds. A NLO generation of the full process,
though possible, is computationally too demanding for our purpose.
We therefore adopt approximations involving the presence
of either an intermediate top quark or a $W$ boson. However, we should
ensure that we do not lose any information about
spin correlations. We thus generate the following samples. 
\begin{itemize}
\item The full matrix element up to the leptons $(b\nu l j)$ in {\sc MG5\_aMC} 
(\textit{fullchain}).
\item $Wbj$ production in {\sc MG5\_aMC} and decay the $W$ in {\sc MadSpin}
(\textit{halfchain}).
\item Single top production $(tj)$ in {\sc MG5\_aMC} and decay the top and $W$ in {\sc MadSpin}
(\textit{nochain}). 
\end{itemize}
We have investigated the differences between the three methods at LO,
where all are straightforward to implement. In particular, given that we wish to retain 
spin correlations in all three approaches, we examine the differences
involving the polarisation angle $\theta_i^z$, the angle
between the direction of decay product $i$ and the spectator jet, as
viewed in the top rest frame. The angular distribution of any top
decay product in this frame can be parameterised as
\begin{equation}
  \frac{1}{\sigma}\frac{d\sigma}{d \cos\theta^z_i}=\frac{1}{2}\left(1+a_i P\, \cos\theta^z_i \right)
  \label{eq:pol_angle} 
\end{equation}
where $P$ denotes the top quark polarisation and $a_i$ encodes how
much spin information is transferred to each decay product. For the
charged lepton $a_l$ is close to 1, indicating nearly 100\%
correlation. 
\begin{figure}[h]
\centering
\includegraphics[scale=0.6,page=7]{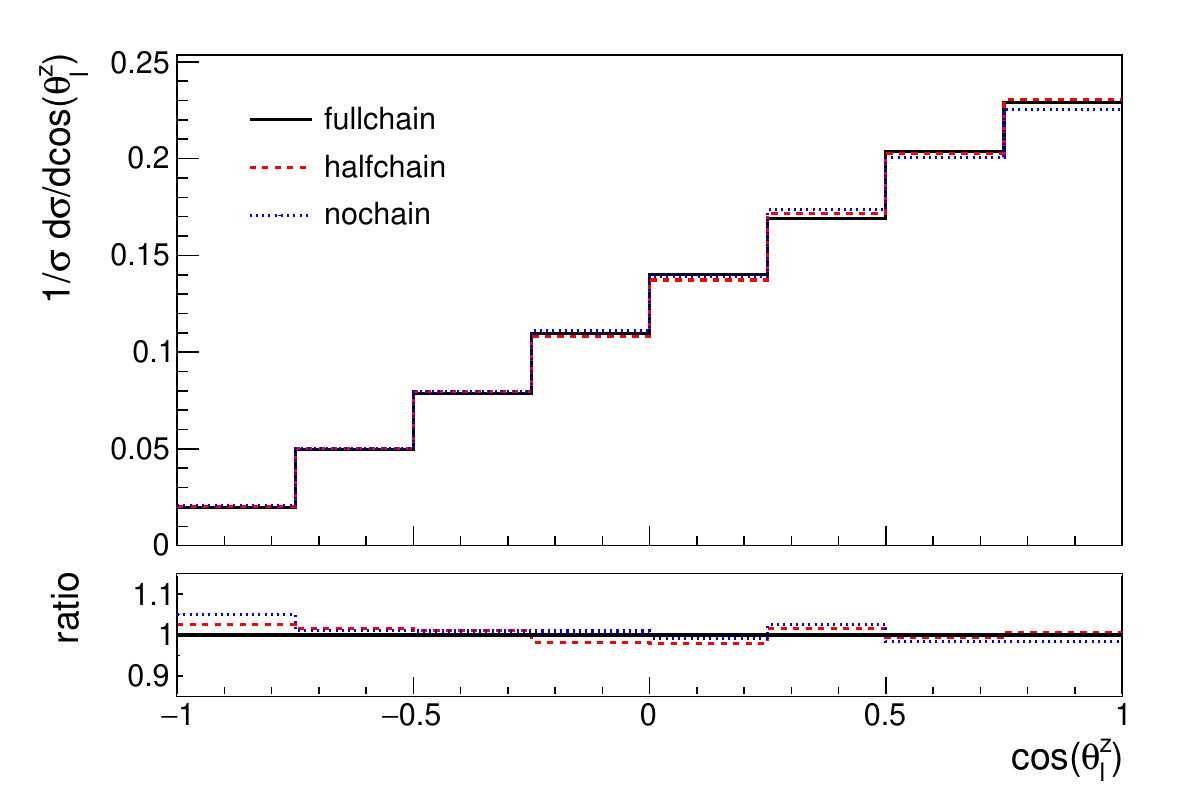}
\caption{The SM top polarisation angle at LO for the 3 different generation options, as described in the text. The ratios with respect to the fullchain method are shown in the lower pane.}
\label{fig:toppol_chain}
\end{figure}
All three options show good agreement, as shown in
fig.~\ref{fig:toppol_chain}. We verified this to be the case for other observables as well. Given the level of agreement we find at
LO between the $Wbj$ and $l\nu b j$ distributions we will follow the
\textit{halfchain} method for our NLO results, i.e. we produce $Wbj$
and decay the $W$ in {\sc MadSpin}, employing the relatively narrow $W$-width. 
A similar agreement is expected to hold at NLO, in particular as the leptonic decay of the $W$ is not
sensitive to higher order QCD corrections.


\subsection{Treatment of top quark width and impact of multiple operator insertions}
\label{sec:treatment-top-quark}

\begin{figure}[H]
\centering
\includegraphics[scale=0.5,page=2]{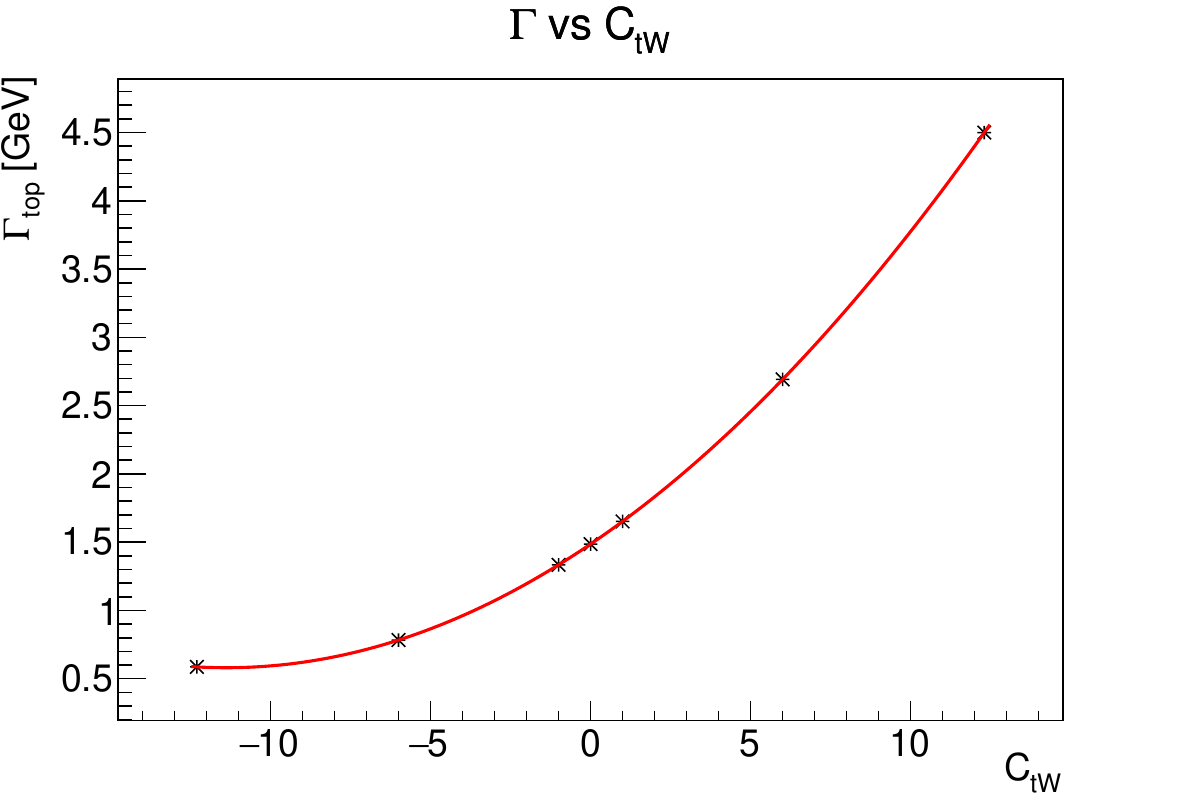}
\caption{The top width as a function of the effective coupling $C_{tW}$
  for $\Lambda=1$ TeV. The quadratic dependence on $C_{tW}$ is
  predicted by equation \ref{eq:width_pol2}. At the indicated points 
the width is computed for the corresponding $C_{tW}$ values, while the line
 is a quadratic fit. }
\label{fig:EFT1_quadratic}
\end{figure}

In addition to the difficulties already present in the
SM calculation for single top production and decay at NLO, the
following EFT related subtleties affect the computation as well:
\begin{itemize}
\item[i)] The width of the top enters in the production of the $Wbj$
  final state. The effective operators affect the numerical value of
  this width, which has to be computed accordingly. We 
  examine the modifications of the width value and its impact on the validity of the
  narrow width approximation for the top decay.
\item[ii)] By considering the $Wbj$ production matrix elements, the
  effective operators can now enter both in top production and in
  top decay. Allowing more insertions in the amplitude
  generates higher order terms in $1/\Lambda^2$. These higher-order terms
  are expected to be suppressed but we will check this explicitly.
  Studying the $W b j$ final state moreover implies that configurations without top
  quarks contribute. The dimension-6 operators can affect also these irreducible
  backgrounds, hence their contributions should be included and their impact studied.
\end{itemize}
Let us address these two subtleties in turn.

i)\, As discussed in equation \eqref{eq:m_squared}, the effect of one
effective operator on the width of the top can be described by a
second order polynomial $1/\Lambda^2$, e.g. for $O_{tW}$ (real $C_{tW}$) the width takes the form:
\begin{equation}
  \Gamma_{\text{top}}(C_{tW}) = \Gamma_{\text{SM}} + \frac{1{\rm TeV}^2}{\Lambda^2} C_{tW} \, \Gamma_{tW} + \frac{1{\rm TeV}^4}{\Lambda^4} C^2_{tW} \, \Gamma_{tW,tW}.
  \label{eq:width_pol2}
\end{equation}
In fig.~\ref{fig:EFT1_quadratic} we show how the top width, computed
at LO, varies as a function of the operator coefficient $C_{tW}$,
demonstrating the quadratic functional dependence. It is important to
stress here that there are experimental constraints on the value of
the width by both CMS and ATLAS
\cite{Aaboud:2017uqq,CMS-PAS-TOP-16-019}, as well as theoretical
proposals \cite{Giardino:2017hva} to extract more information about the
top width.
 
When the width is small compared to the total mass of the
particle, one can factorise the total cross section for a given decay
channel into the production cross section multiplied by the branching
ratio corresponding to that particular decay channel. This narrow
width approximation (NWA) rests upon the following approximation for the 
denominator of the squared top quark propagator \cite{NWA}:
\begin{equation}
\frac{1}{(p^2-M_{\text{top}}^2)^2+M_{\text{top}}^2\Gamma_{\text{top}}^2} \xrightarrow{\sfrac{\Gamma_{\text{top}}}{M_{\text{top}}} \rightarrow 0} \frac{\pi}{M_{\text{top}}\Gamma_{\text{top}}}\delta(p^2-M_{\text{top}}^2).
\label{eq:nwa}
\end{equation}  
The inclusive cross section of the single top production and decay to a $W$ boson and a $b$ quark is then approximated by:
\begin{equation}
\sigma(pp \rightarrow Wbj) \rightarrow \sigma(pp \rightarrow tj) \, \frac{\Gamma(t \rightarrow Wb)}{\Gamma_{\text{top}}} = \sigma(pp \rightarrow tj) \,  \text{BR}(t \rightarrow Wb).
\label{NWAwidth}
\end{equation}

\begin{figure}[h]
\centering
\includegraphics[scale=0.5,page=1]{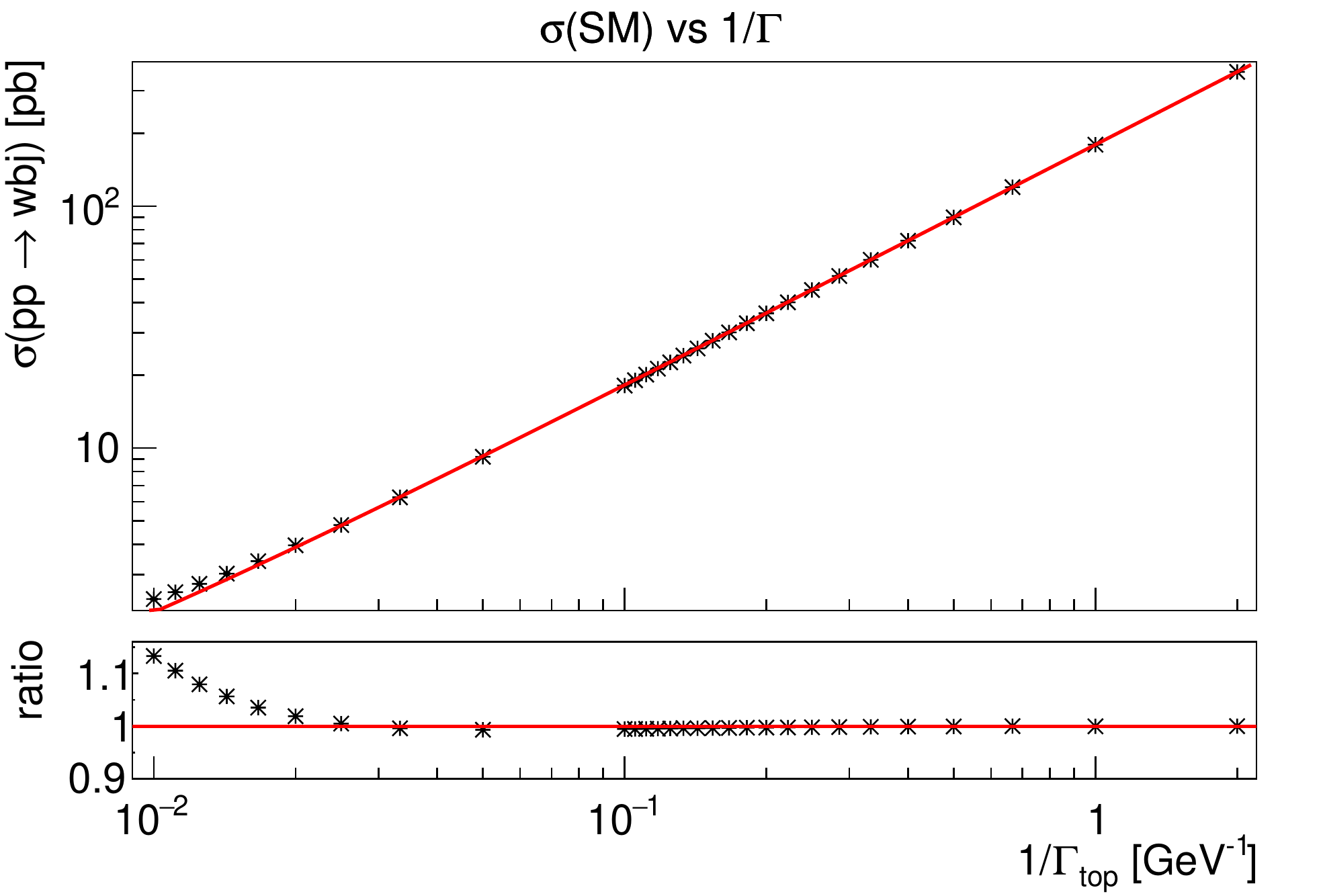}
\caption{The SM cross section as a function of the width of the top. The NWA is valid when the relation is linear.}
\label{fig:cross_vs_width}
\end{figure}

Since for top decays the branching ratio $\text{BR}(t \rightarrow Wb) \approx 1$, a direct
way of testing the range of NWA validity in (\ref{NWAwidth}) is to calculate $\sigma(pp
\rightarrow Wbj)$ at different numerical values of $\Gamma_{\text{top}}$, 
with SM couplings. This is shown
in fig.~\ref{fig:cross_vs_width} where the linear dependence on $1/\Gamma_{\text{top}}$ can be observed for small $\Gamma_{\text{top}}$,  
whilst for $\Gamma_{\text{top}}> 50$ GeV the linear dependence breaks down. 
For non-excluded values of the operator coefficients the modifications of the width are moderate 
and therefore the NWA is expected to hold. 

\begin{figure}[h]
  \centering
  \includegraphics[scale=0.5,page=5]{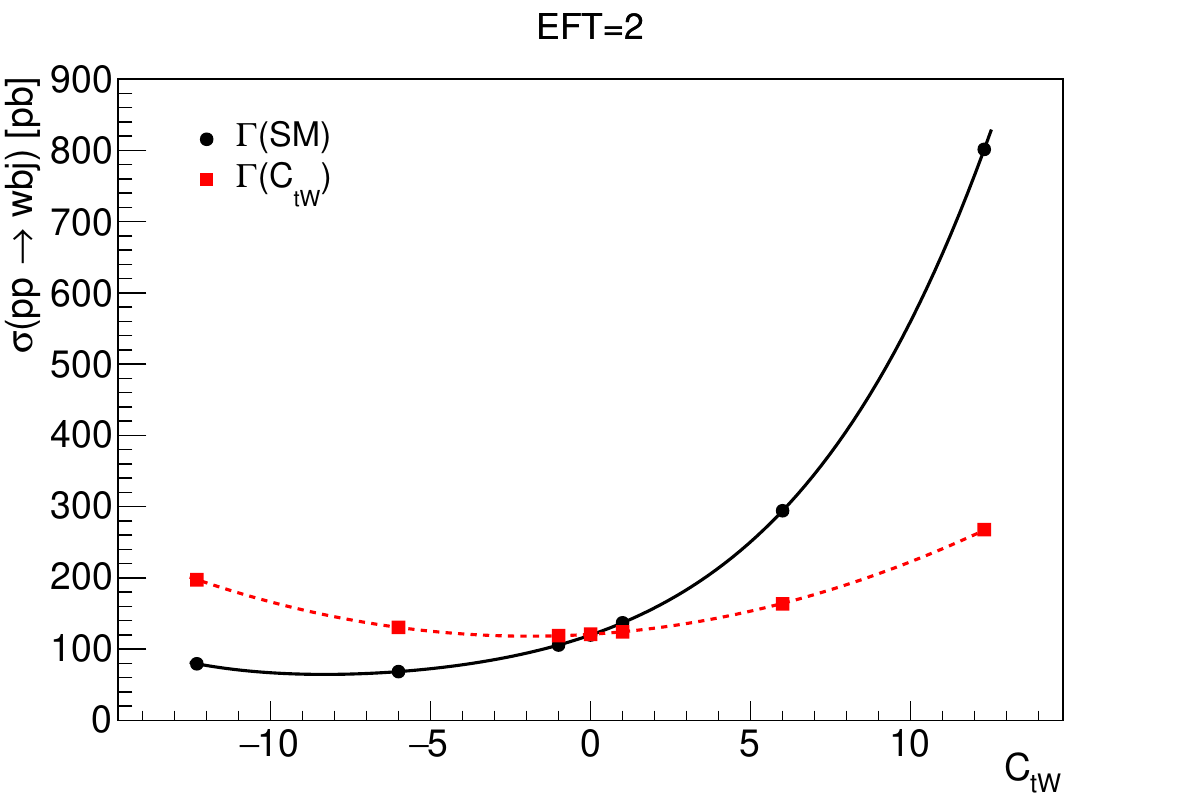}
  \caption{The cross section of the $Wbj$ production as a function of
    $C_{tW}$ with two EFT insertions for the width of the top fixed to
    the SM value of 1.5 GeV (quartic dependence), or computed according to the value of the
    operator (quadratic dependence). }
  \label{fig:EFT2}
\end{figure}

ii) The same interactions occur, at leading order, in the production and the decay of the single top quark,
hence the amplitudes for the process $\sigma(pp \rightarrow Wbj)$ can
contain up to two insertions of an effective operator (to be precise of $O_{tW}$ or $O_{\varphi Q}^{(3)}$). The behaviour of
the cross section as a function of the coefficient requires then a
more complicated functional form than the one predicted by 
(\ref{eq:xsecpara}), in part due to the presence of more
insertions, and in part due to the dependence of the top width on the coefficient, which enters in the $Wbj$ calculation.  The
situation is however simplified in the NWA since the cross section for
the production and the decay of a single top quark with two insertions
of the effective coupling $C_{tW}$ can then be written schematically as:
\begin{equation}
  \sigma^{pp \rightarrow Wbj}_{\text{EFT=2}}(C_{tW},\Gamma(C_{tW}))
  \sim \left( \sigma_{\text{SM}} 
+ C_{tW} \cdot \sigma_{tW} + C^2_{tW} \cdot \sigma_{tW,tW} \right)_{(\text{tj})}\,,
  \label{eq:EFT2_full}
\end{equation}
where we have chosen $\Lambda = 1\, \mathrm{TeV}$ to avoid notational
clutter. We shall also do this for eq.~(\ref{eq:EFT1_full}) below.
We have indicated by the subscript $(tj)$
that the dependence of the partial $Wb$ width and total width on $C_{tW}$
(and therefore in the branching fraction) 
cancels in equation \eqref{NWAwidth}.   In other words, in the NWA the $C_{tW}$
dependence is as for producing a stable top quark plus jet. Fig.~\ref{fig:EFT2} compares
the case where the width is fixed to its SM value (1.5 GeV) with the
case where the width is computed based on the coefficient value. In
both cases two insertions are allowed in the amplitude. 
When working in the NWA, the width, being a function of the
coefficient, eventually leads to a quadratic dependence of the 
cross section on $C_{tW}$ in \eqref{eq:EFT2_full}. 
When one takes the width fixed there is no cancellation
in the partial and total top width, and the dependence is quartic.

\begin{figure}[h]
\centering
\includegraphics[scale=0.5,page=4]{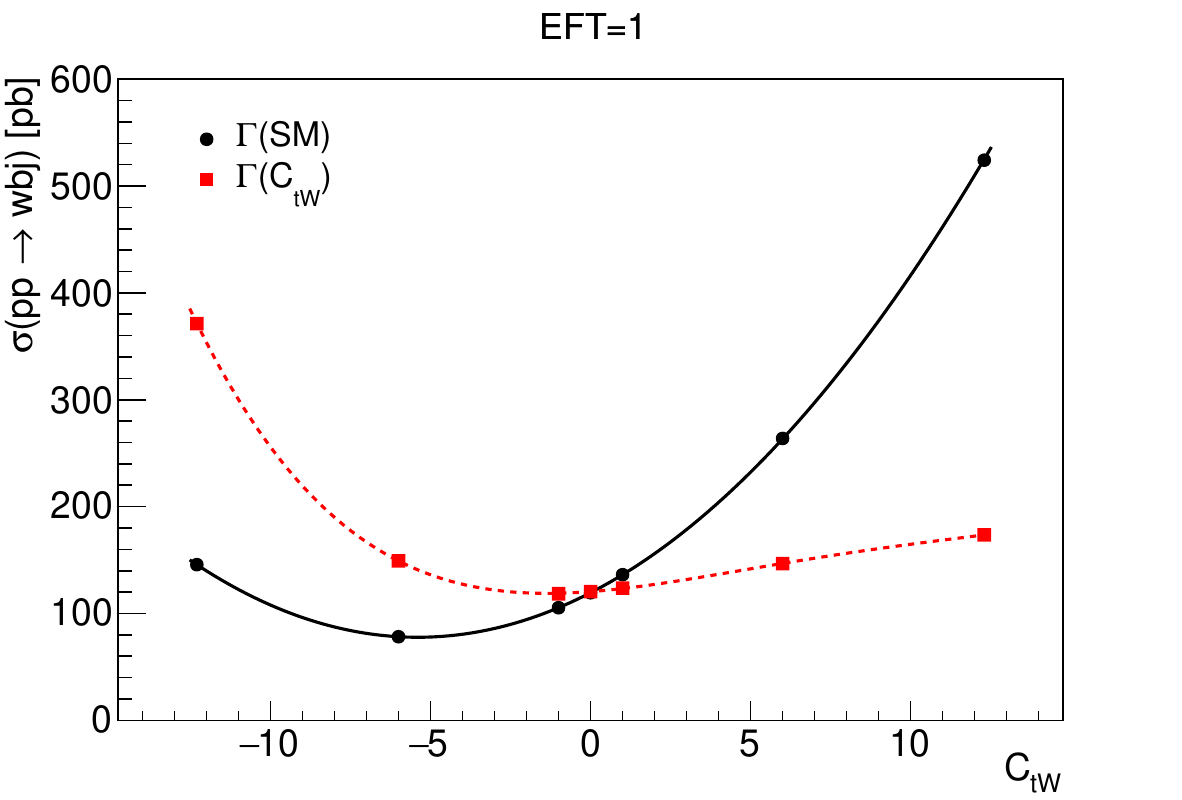}
\caption{The cross section of $Wbj$ production as a function of
  $C_{tW}$ with one insertion with the width of the top fixed to the
  SM value (quadratic behaviour) or computed according to the value of
  the operator (higher-order polynomial).}
\label{fig:EFT1}
\end{figure}

When only one insertion of an effective coupling is
allowed (still in the NWA), it can enter either in the production or in the decay. The
simplified form of the cross section in this case becomes:
\begin{equation}
 \sigma^{pp \rightarrow Wbj}_{\text{EFT=1}}(C_{tW}) 
\sim \frac{ \sigma_\text{SM} + C_{tW} \, \sigma_{tW} + C^2_{tW} \,
    \sigma_{tW,tW}}{\Gamma_{\text{SM}} + C_{tW} \, \Gamma_{tW}  + C^2_{tW} \, \Gamma_{tW,tW}}
\label{eq:EFT1_full}
\end{equation}
where $\sigma$ indicates that the $Wbj$ final state is generated, with
only one operator insertion. The $\Gamma$ in the
denominator indicates that the cross section is described by the
narrow width approximation. Since the terms in the numerator are
different in their $1/\Lambda^2$ dependence from the terms in the denominator, no cancellations
occur. The impact of how the width is treated can be seen in fig.~\ref{fig:EFT1} for
the one-insertion calculations, where as expected a quadratic behaviour
is observed when the width of the top is fixed, and a higher order
polynomial is required to describe the behaviour when the width is
computed with $C_{tW}$ dependence.

\begin{figure}[h]
\centering
\includegraphics[scale=0.6,page=3]{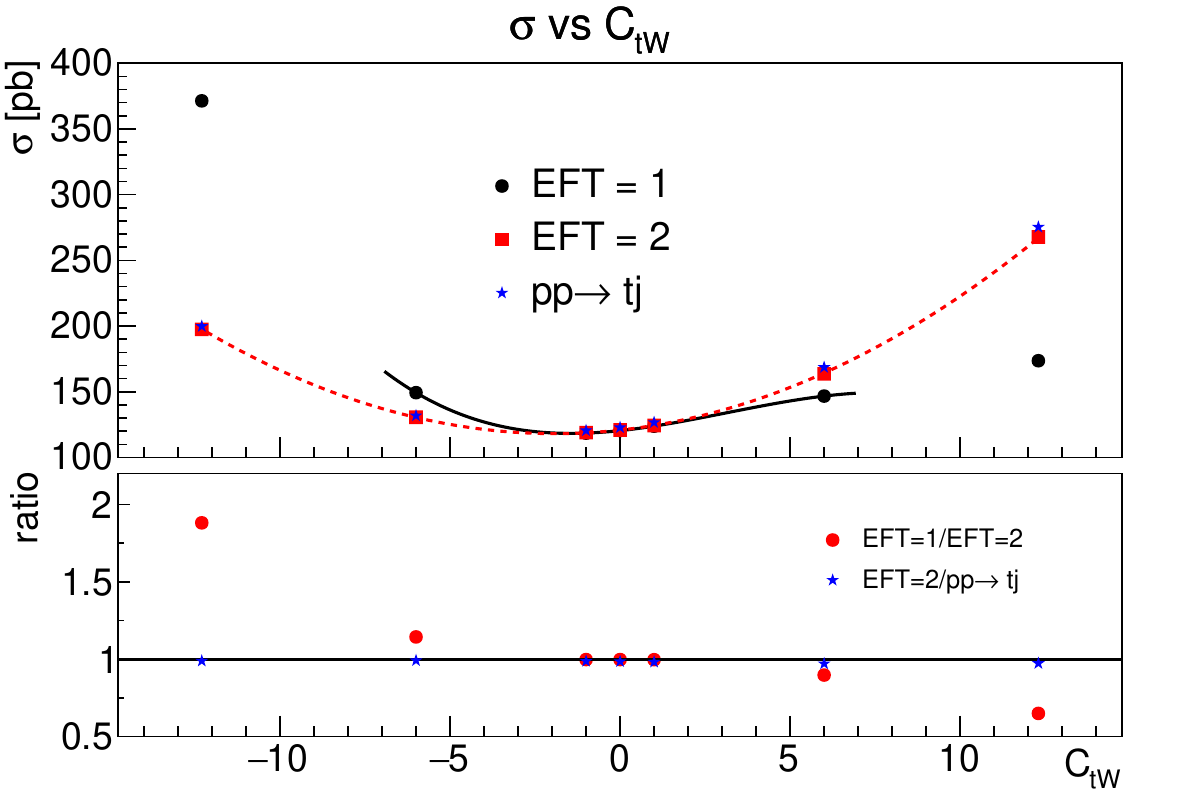}
\caption{Comparing the different behaviour of the $Wbj$ cross section with
  one $C_{tW}$ insertion (EFT=1) or two $C_{tW}$ insertions
  (EFT=2). Both effects have been discussed separately in
  figs. \ref{fig:EFT2} and \ref{fig:EFT1}. Additionally we show the
  production cross section ($pp \rightarrow tj$) which is reproduced
  by $Wbj$ with two insertions when the right width is taken into account.}
\label{fig:cross_vs_ctW}
\end{figure}

Finally in fig.~\ref{fig:cross_vs_ctW} we compare the behaviour of the total
cross section with one operator insertion (EFT=1) or two insertions
(EFT=2). It can be observed that for small values of the coupling, the
linear term dominates and the cross sections coincide, as they only
differ by higher order terms in $1/\Lambda^2$.
Notice that fig.~\ref{fig:cross_vs_ctW} also shows that the
production cross section ($\sigma(pp \rightarrow tj)$) is very close to the
$Wbj$ cross section with two insertions of the couplings, as we expect
from the NWA.



\begin{figure}[H]
\centering
\begin{subfigure}{0.5\textwidth}
\centering
\includegraphics[scale=0.4,page=1]{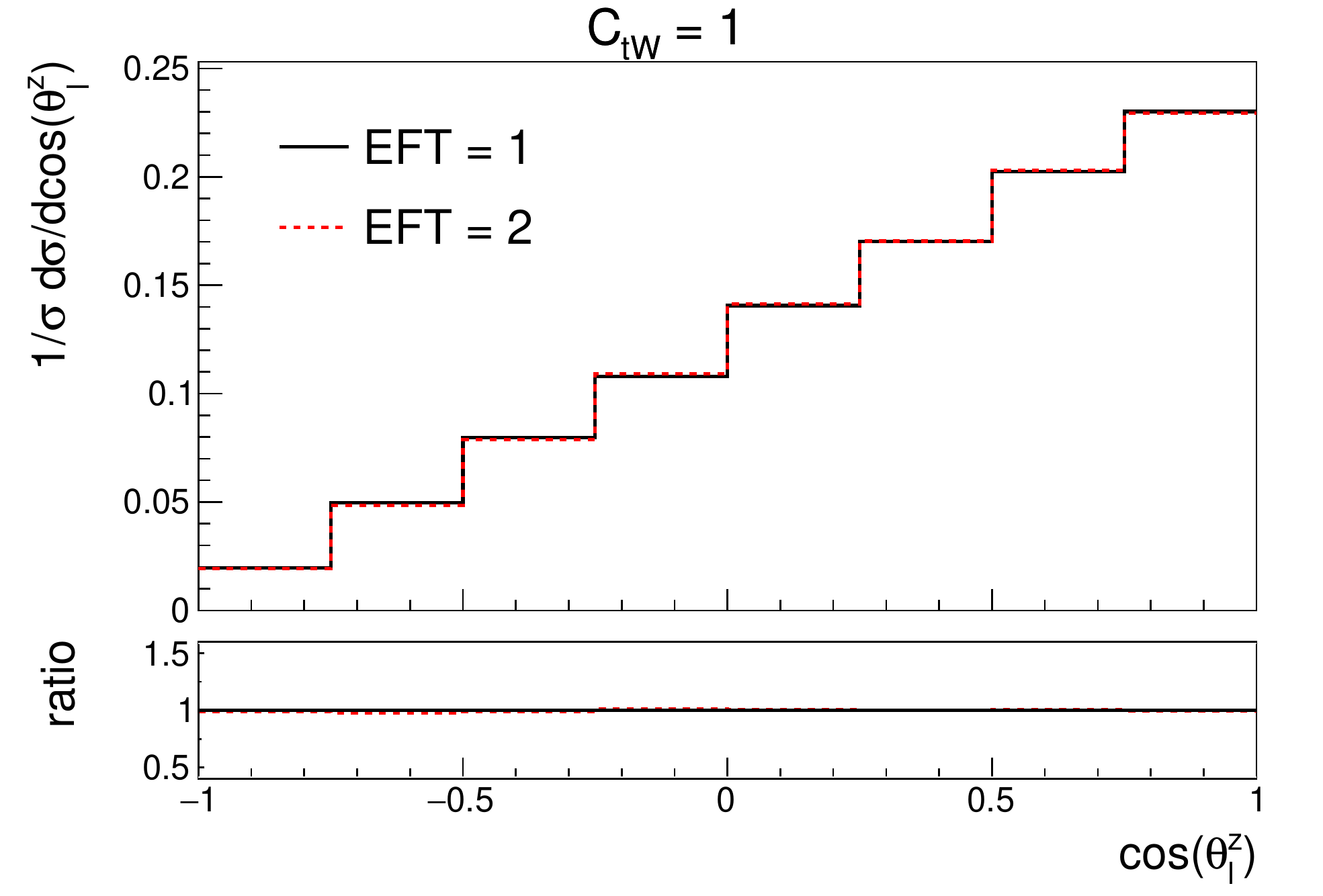}
\end{subfigure}%
\begin{subfigure}{0.5\textwidth}
\centering
\includegraphics[scale=0.4,page=2]{LO_Width_effects_toppol.pdf}
\end{subfigure}
\caption{The top polarisation angle at LO with different values for
  the $O_{tW}$ effective operator. On the left hand side results for
  $C_{tW}=1$ are shown, whilst $C_{tW}=6$ on the right hand side. The
  comparison between one and two EFT insertions is shown.}
\label{fig:EFT1_vs_EFT2_toppol}
\end{figure}

In order to examine whether the conclusions reached so far apply to differential
distributions as well we show in fig.~\ref{fig:EFT1_vs_EFT2_toppol} the top
polarisation angle, defined in (\ref{eq:pol_angle}), obtained for two different values of the 
coefficient for one and two EFT insertions. The left pane shows both
EFT options for $C_{tW}=1$. One can observe that the two distributions
coincide within statistical errors. The right pane shows the case of 
$C_{tW}=6$, here the impact of higher order terms are important and
these cannot be described by a global normalisation factor as shown in the ratio inset. This
indicates that higher order effects in the EFT can be non-negligible. 
Therefore, for consistency with the production cross section and to
avoid missing large higher order effects, all distributions in the
rest of this paper have been obtained by generating $Wbj$ allowing up to two EFT
insertions, with the top width computed as a function of the coupling.

We note here that we validated our leading-order results with the ones
discussed in \cite{Aguilar-Saavedra:2014eqa} for the top-quark
polarisation ($P$), analysing power ($a_i$) and lepton angular
distributions. We performed a detailed comparison by allowing all
possible insertions of the operators and matching all parameters of
the computation with the one implemented in the generator {\sc Protos} \cite{AguilarSaavedra:2008gt},
and found perfect agreement.

\subsection{Results at NLO}

Having studied the various effects at LO we proceed by computing the
$Wbj$ cross section at NLO in QCD in the presence of the dimension-6
operators. The $W$ boson is decayed leptonically through {\sc MadSpin},
and {\sc Pythia8} \cite{Sjostrand:2014zea} is used for parton
showering and hadronisation. Since we also generate the irreducible
backgrounds, a loose invariant mass cut is imposed on the $Wb$ system,
centered on the top mass 
$100\, \mathrm{GeV} < M_{Wb-\mathrm{jet}} < 250\, \mathrm{GeV}$ 
\cite{Frederix:2016rdc}. Jet clustering is done using {\sc
  fastjet} \cite{Cacciari:2011ma} and the anti-$k_t$ algorithm
\cite{Cacciari:2008gp}, with the jet radius parameter set to 0.4. 
All other generator settings and kinematic cuts are
 the same as in section \ref{sec:tj}.\\
 
 \begin{figure}[h]
\centering
\begin{subfigure}{0.5\textwidth}
\centering
\includegraphics[scale=0.4,page=1]{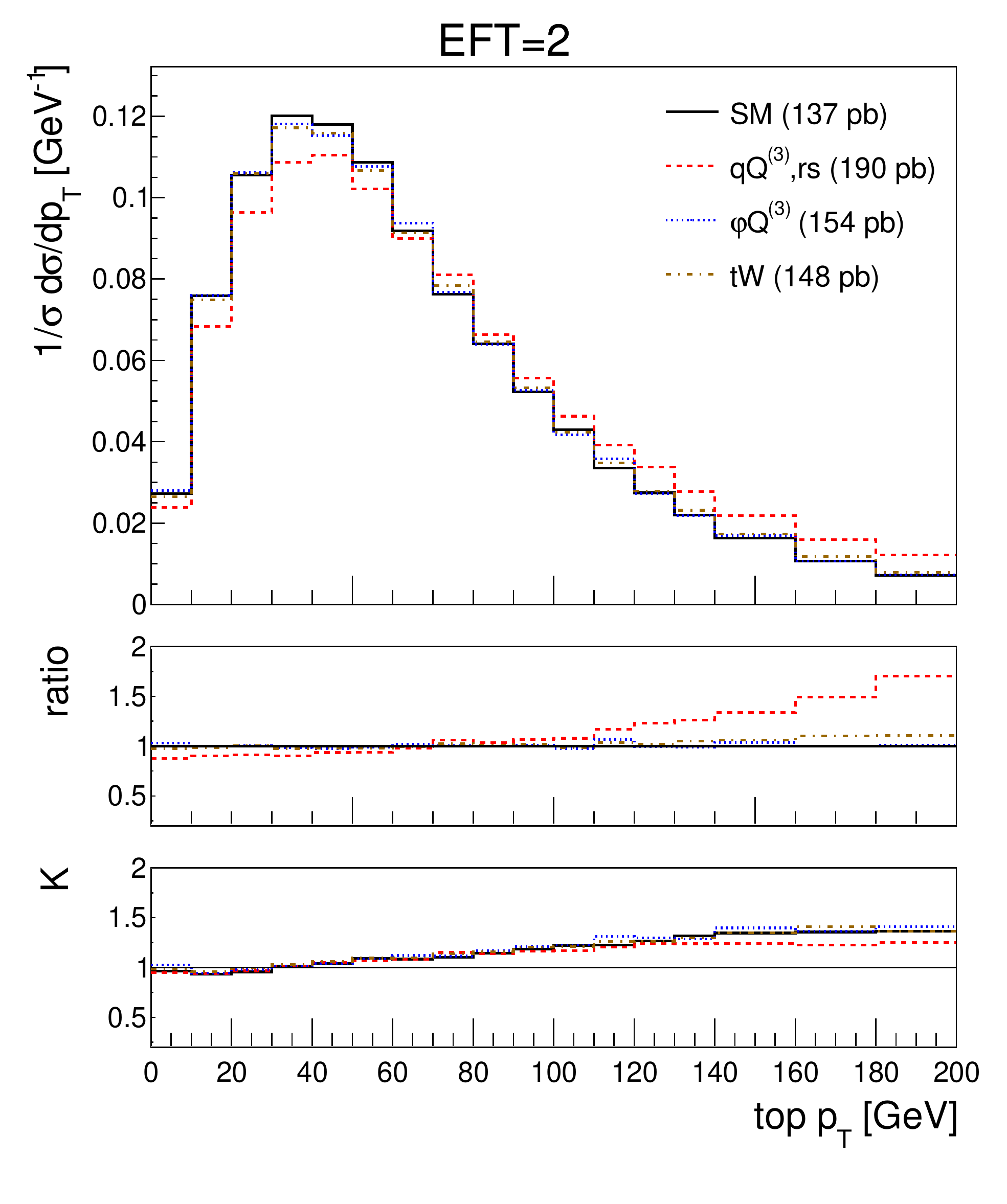}
\end{subfigure}%
\begin{subfigure}{0.5\textwidth}
\centering
\includegraphics[scale=0.4,page=2]{EFT2_plots_baseline.pdf}
\end{subfigure}
\caption{The NLO distributions of the reconstructed top transverse momentum (left)
  and rapidity (right) for the SM and the three effective operators of
  interest for the couplings values of table \ref{tab:cvalues}. 
We quote in the left figure the corresponding inclusive cross section from
this table.
The ratio shown in the first inset is defined as the
  effect of the operator over the SM, the second inset shows the
  $K$-factor.}
\label{fig:EFT2_pt+eta}
\end{figure}

We start by showing the top quark transverse momentum
and rapidity distributions in fig. ~\ref{fig:EFT2_pt+eta} for the SM and the three operators, along with the
ratio over the SM prediction and the corresponding $K$-factor. 
The top quark is now reconstructed from its semi-leptonic decay products, consisting of
hardest electron, the associated neutrino and a $b$-jet. The light spectator jet is identified as 
well. When more than one $b$-jet is present we choose
the one yielding the best reconstructed top mass. The results in
fig.~\ref{fig:EFT2_pt+eta} are in excellent agreement with those in fig.~\ref{fig:production_pt+eta}.\\

\begin{figure}[h]
\centering
\begin{subfigure}{0.5\textwidth}
\centering
\includegraphics[scale=0.4,page=3]{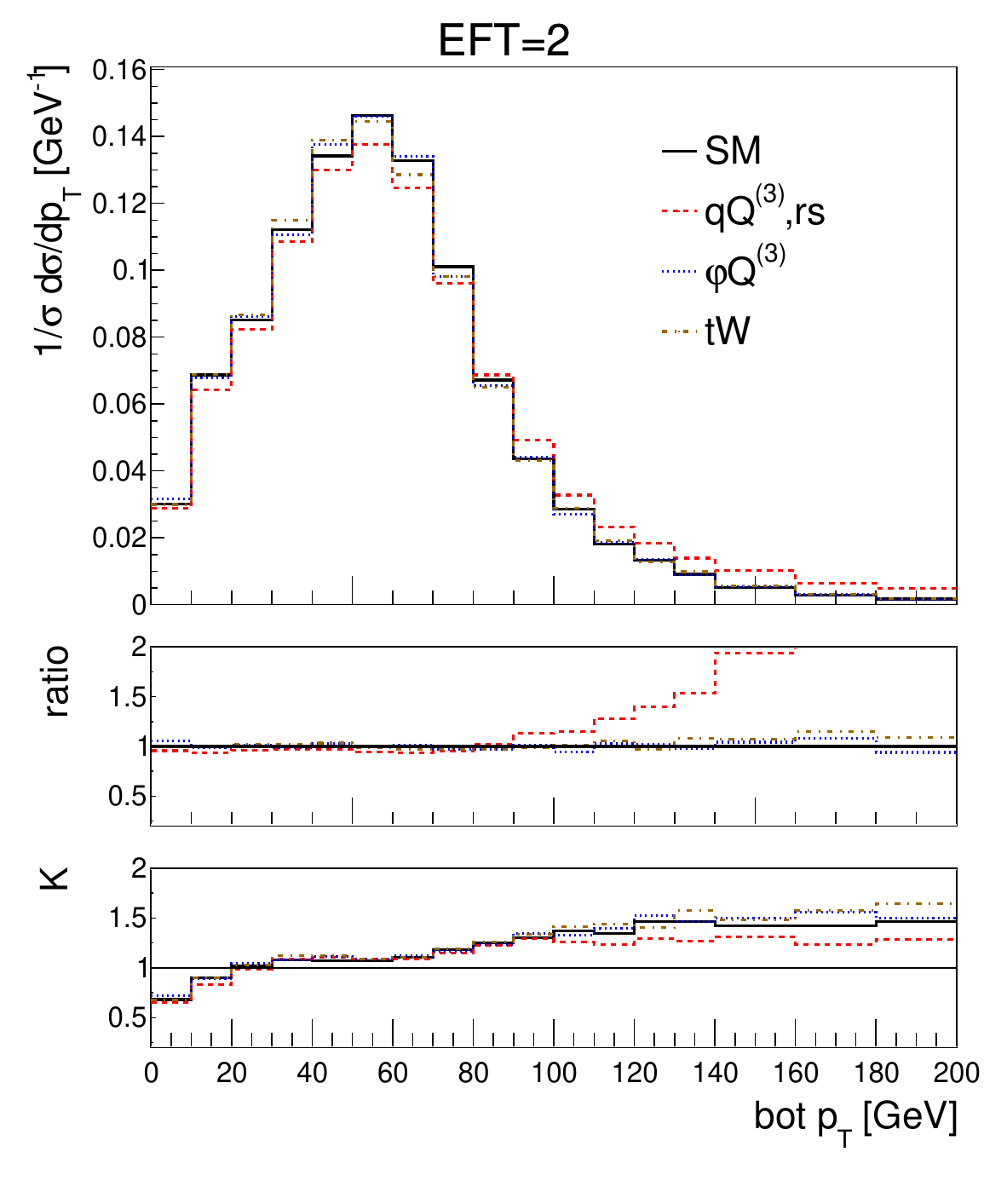}
\end{subfigure}%
\begin{subfigure}{0.5\textwidth}
\centering
\includegraphics[scale=0.4,page=7]{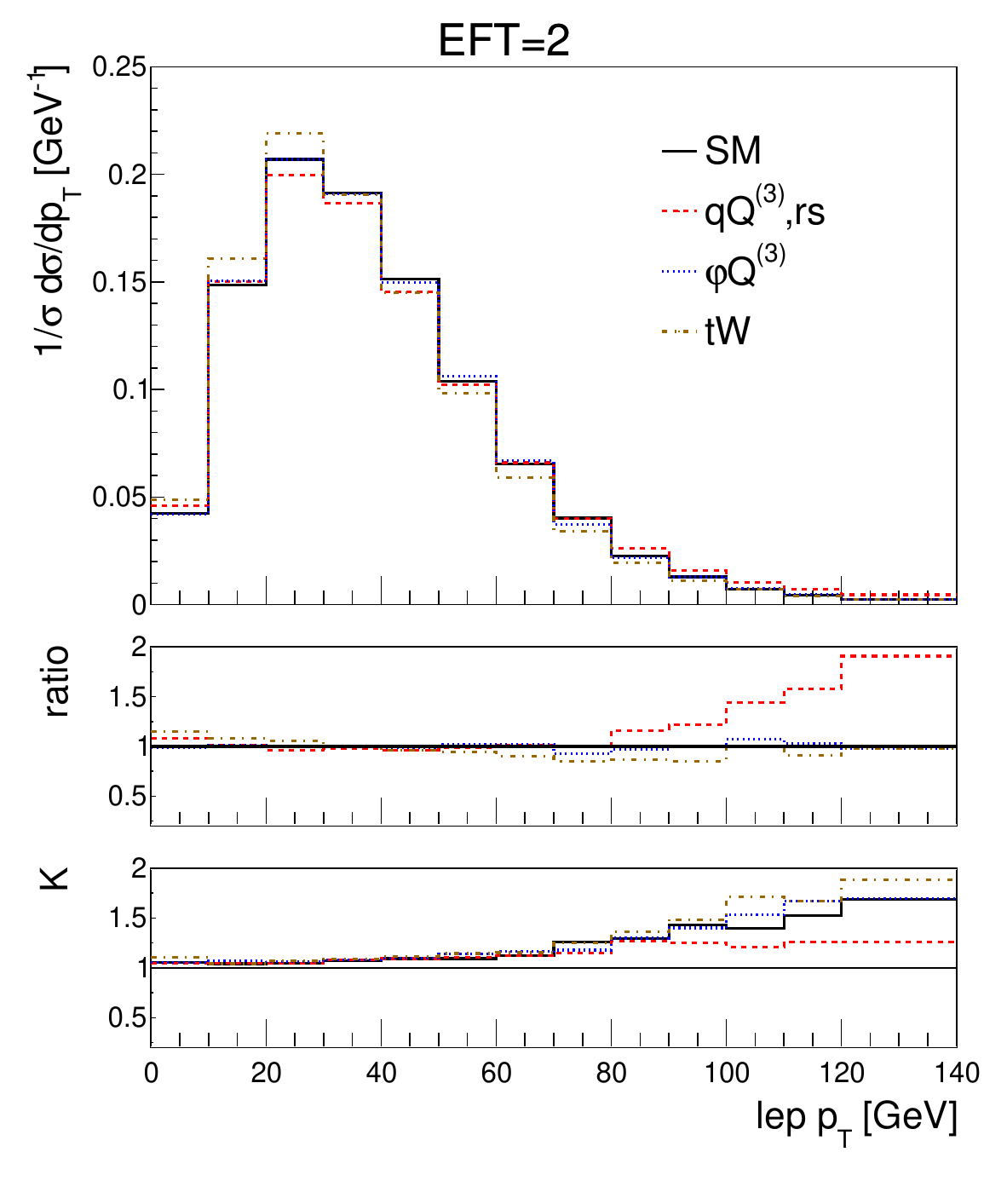}
\end{subfigure}
\caption{The NLO distributions of the $b-$jet (left) and
  electron (right) transverse momentum for the SM and the three
  effective operators of interest for the couplings values of table \ref{tab:cvalues}. 
  The ratio shown in the first inset is defined as the
  effect of the operator over the SM, the second inset shows the
  $K$-factor.}
\label{fig:EFT2_pt_b_lep}
\end{figure}

Other observables of interest are the kinematic distributions of the lepton and
$b$-jet from the decay of the top, shown in
fig.~\ref{fig:EFT2_pt_b_lep}. Their $p_T$ distributions show a harder tail
for the 4-fermion operator, whilst all contributions show a non-flat $K$-factor, with QCD corrections being larger in the high-$p_T$ region,
for both the $b$-jet and the lepton.\\

Since the spin axis of the top is known \cite{Mahlon:1996pn} a rich
set of angular observables showing spin correlations, can be
exploited. Below we will elaborate on the definitions of the angles
involved. 
In general, based on the choice of reference frame, it is possible to probe the production- and decay vertex of
the single top separately. In any
frame, 
a new set of coordinates can be defined based on the spin axis of the
top. 
These additional coordinate axes provide the ability to construct
other angles that contain spin information. For brevity, only the
angular distributions that show the most sensitivity to the effective
operators will be presented in this section.
      

The polarisation angle defined in equation \eqref{eq:pol_angle} is one of the spin correlated angles that probes the production vertex. We use the same reference system as in \cite{Aguilar-Saavedra:2014eqa} to construct a new set of coordinates:
\begin{equation}
  \hat{z} = \frac{\vec{p_j}}{|\vec{p_j}|}, \;\;\;\;\;\; \hat{y} = \frac{\vec{p_j} \times \vec{p_q}}{|\vec{p_j} \times \vec{p_q}|}, \;\;\;\;\;\; \hat{x} = \hat{y} \times \hat{z}\,.
\label{eq:pol-3D}
\end{equation}
The vectors $\vec{p}_j$ and $\vec{p}_q$ represent the direction of the spectator- and of the
initial quark, respectively,  both in the top quark rest-frame.
Since the initial quark cannot be known with certainty, the beam axis is used.\\

\begin{figure}[h]
\centering
\begin{subfigure}{0.5\textwidth}
\centering
\includegraphics[scale=0.35,page=11]{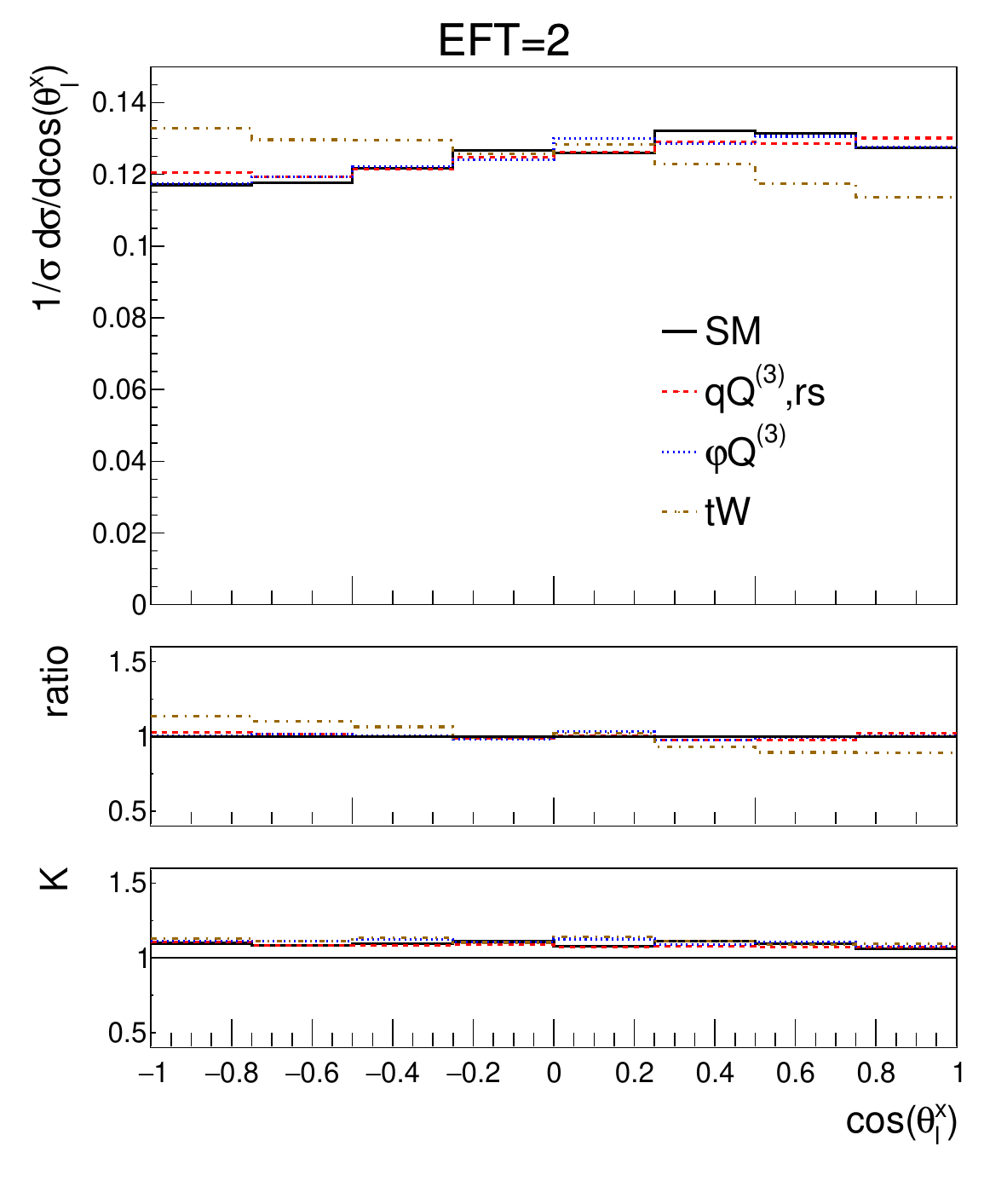}
\end{subfigure}%
\begin{subfigure}{0.5\textwidth}
\centering
\includegraphics[scale=0.35,page=12]{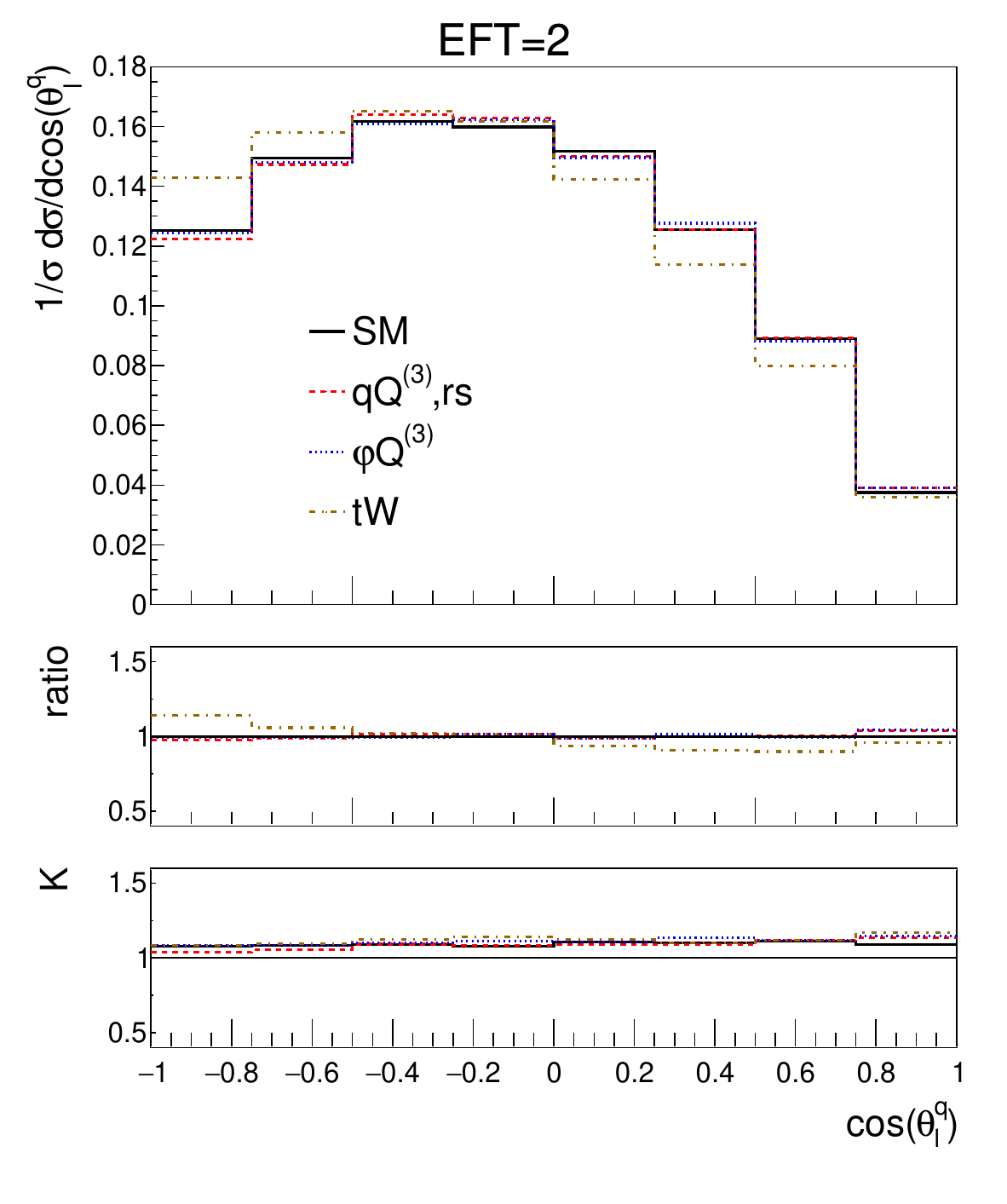}
\end{subfigure}
\caption{The NLO distributions of the top polarisation angle (left)
  and W helicity (right) for the SM and the three effective operators
  of interest for the couplings values of table \ref{tab:cvalues}. The ratio shown in the first inset is defined as the
  effect of the operator over the SM, the second inset shows the
  $K$-factor.}
\label{fig:EFT2_pol+hel}
\end{figure}

We investigate the distributions of the angles between the directions of
the top quark decay products and these new directions. The angle of
the charged lepton with respect to the three axes defined above is affected most by the polarisation of the top \cite{Mahlon:2000ze}.
 Fig.~\ref{fig:EFT2_pol+hel} (left) shows the NLO
distributions for $\cos\theta_l^x$, where $\theta_l^x$ is the angle
between the lepton and direction $\hat{x}$. 
Notice that the dipole operator
($\mathcal{O}_{tW}$) leads to a different distribution compared to the
SM and the other operators. 

In order to probe new interactions in the decay of the top, one can
examine the well-known $W-$helicity fractions $F_+, F_L$ and $F_0$
defined in:
\begin{equation}
  \frac{1}{\Gamma} \frac{d\Gamma}{d \cos \theta_l^q} = \frac{3}{8}
  \left( 1 + \cos \theta_l^q \right)^2 F_+ 
+ \frac{3}{8} \left( 1 - \cos \theta_l^q \right)^2 F_- + \frac{3}{4} \sin^2 \theta_l^q F_0,
\label{eq:Whel}
\end{equation}
where $\theta_l^q$ is the angle between the $W$ in the top rest-frame
and the charged lepton in the $W$ rest-frame. $F_i$ represent the
helicity fractions, with $\sum F_i = 1$. Again a new reference system can be constructed \cite{AguilarSaavedra:2010nx}:
\begin{equation}
  \hat{q} = \frac{\vec{p_W}}{|\vec{p_W}|}, \quad \hat{N} = \frac{\vec{s_t} \times \vec{q}}{|\vec{s_t} \times \vec{q}|}, \quad \hat{T} = \hat{q} \times \hat{N}\,.
\label{eq:Whel3D}
\end{equation}
The vectors $\vec{p}_W$ and $\vec{s}_t$ are both defined in the
rest-frame of the top quark and depict the direction of the $W$ boson
and that of the top quark spin, respectively. The spin of the top
quark is taken as the direction of the spectator jet
\cite{Mahlon:1996pn,Mahlon:1999gz}. The angle of the lepton in the $W$ rest-frame with respect to the three axes defined above probes the decay vertex. Fig.~\ref{fig:EFT2_pol+hel} (right) shows the
NLO distributions for cos\,$\theta_l^{q}$ where the sensitivity to
the dipole interaction comes mainly in the $\theta_l^q\sim\pi$
region. 

\begin{figure}[h]
\centering
\begin{subfigure}{0.5\textwidth}
\centering
\includegraphics[scale=0.35,page=11]{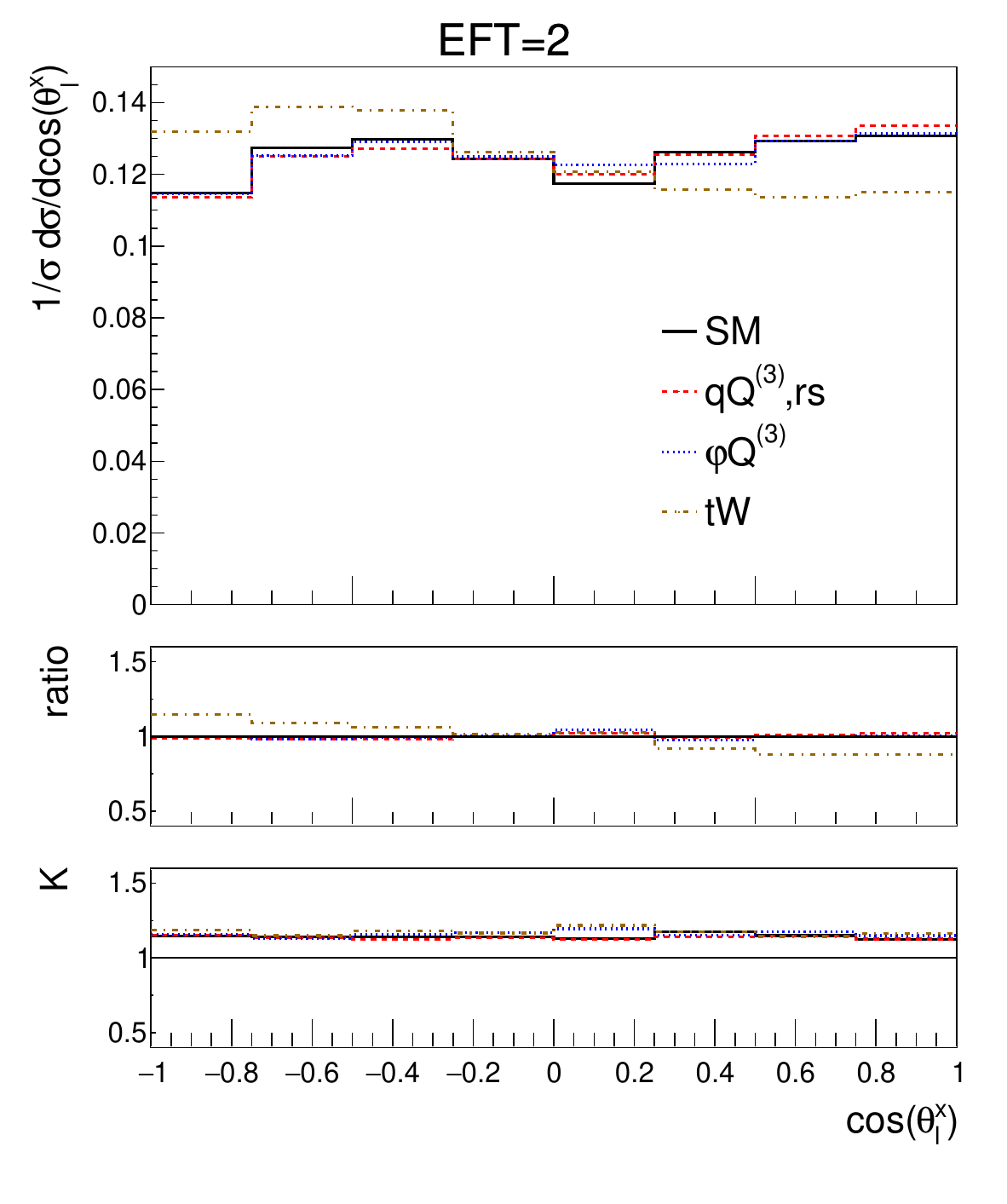}
\end{subfigure}%
\begin{subfigure}{0.5\textwidth}
\centering
\includegraphics[scale=0.35,page=12]{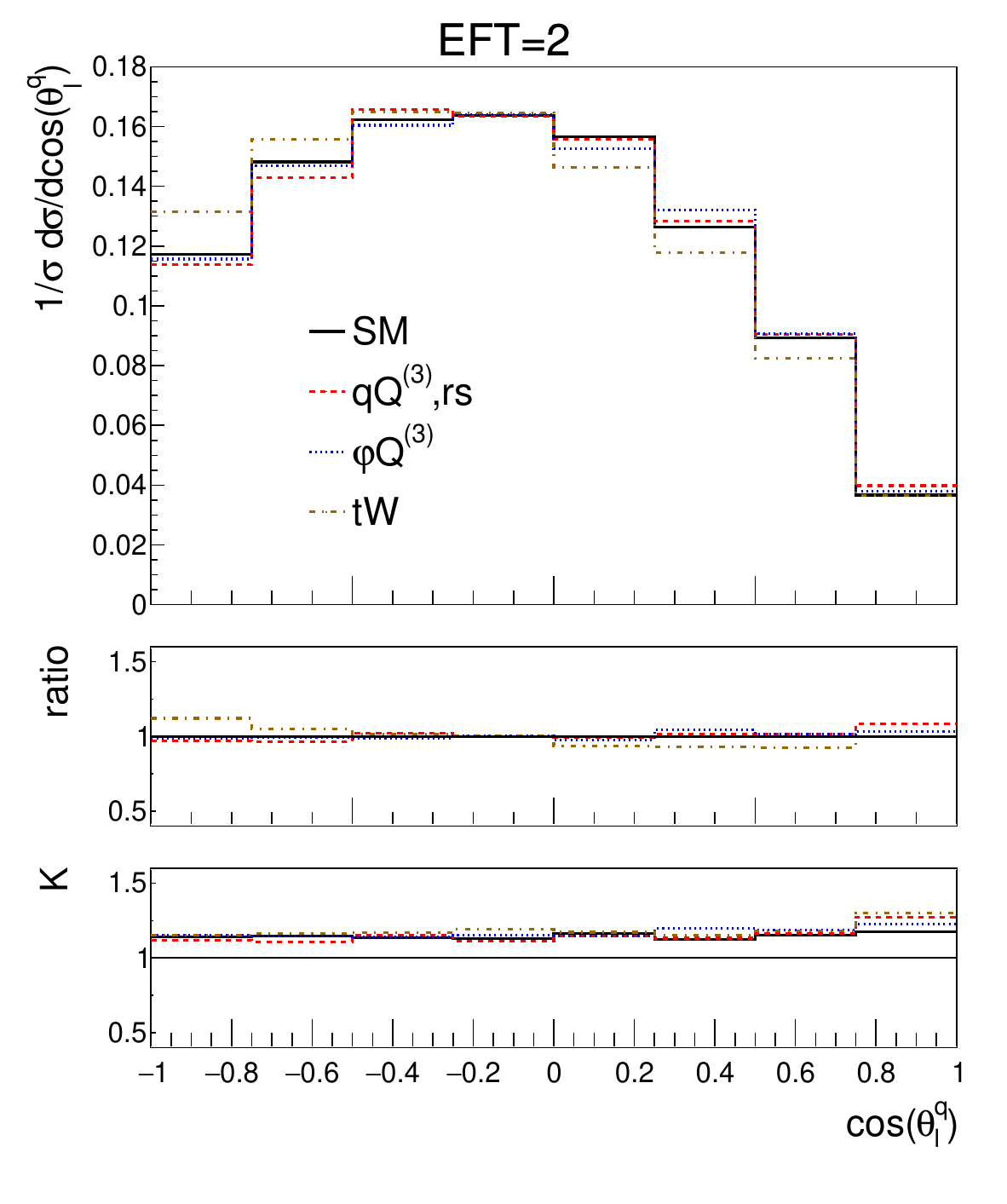}
\end{subfigure}
\caption{The NLO distributions of the top polarisation angle (left)
  and W helicity (right) for the SM and the three effective operators
  of interest for the couplings values of table \ref{tab:cvalues}. The ratio shown in the first inset is defined as the
  effect of the operator over the SM, the second inset shows the
  $K$-factor. Here additional cuts are applied on the leptons: $p_T^l > 10$ GeV and $|\eta^l| < 2.47$ and jets: $p_T^j > 20$ GeV and $|\eta^j| < 4.5$.}
\label{fig:EFT2_pol+hel_cuts}
\end{figure}

\begin{figure}[h]
\centering
\includegraphics[scale=0.35,page=15]{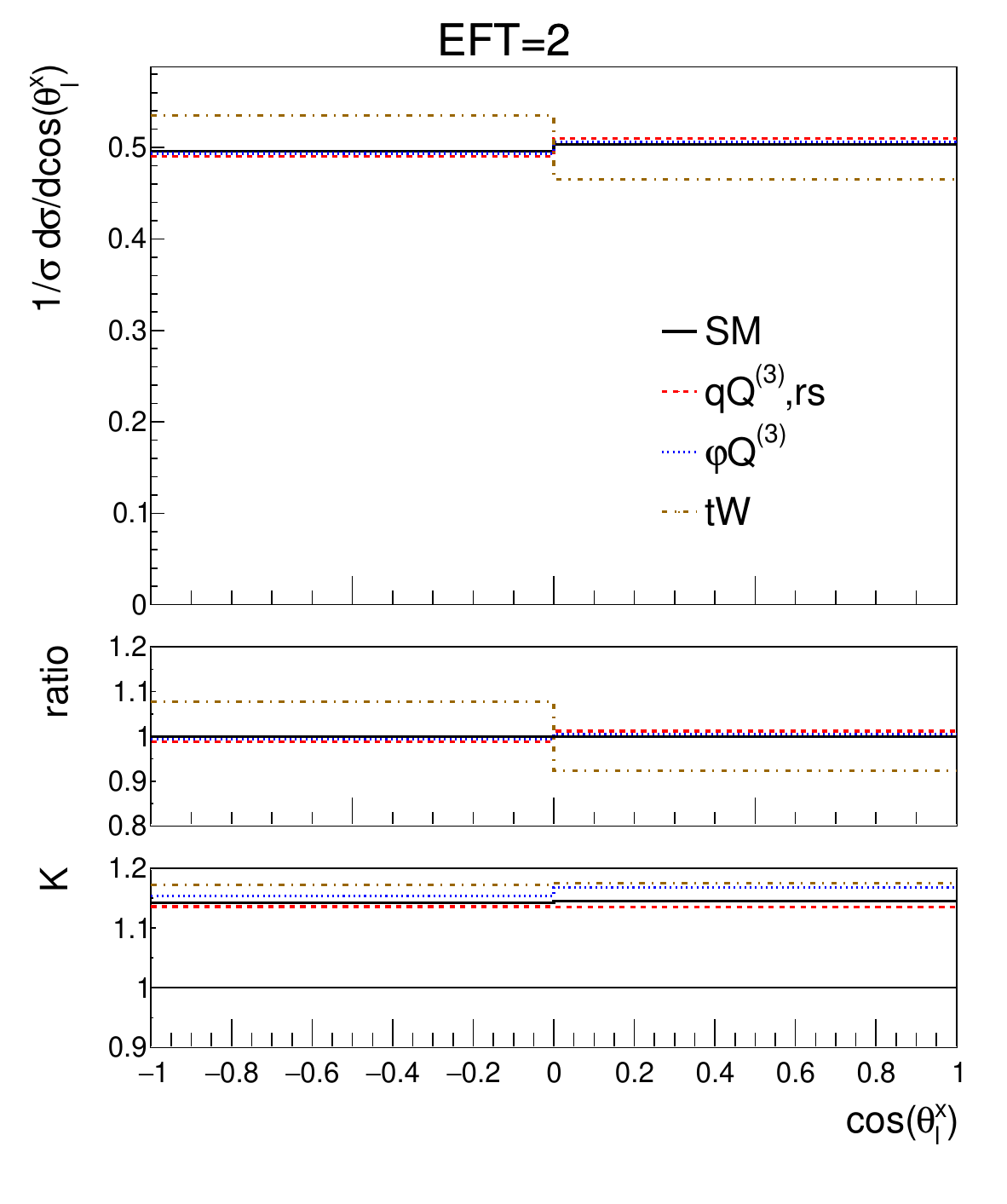}
\caption{The asymmetry between positive and negative values of the top polarisation angle (cos\,$\theta_l^x$) by presenting fig.~\ref{fig:EFT2_pol+hel_cuts} (left) in 2 bins.}
\label{fig:EFT2_pol_2bin}
\end{figure}

To show more realistic distributions, 
fig.~\ref{fig:EFT2_pol+hel_cuts} shows the same observables as 
fig.~\ref{fig:EFT2_pol+hel}, only here additional cuts have been applied resembling the acceptance of the ATLAS detector. Namely, charged leptons must lie inside the $|\eta| < 2.47$  region and have a transverse momentum of at least 10 GeV, whereas jets should have a transverse momentum larger then 20 GeV and lie inside the $|\eta| < 4.5$ region. We note here that
experimental selection cuts can potentially be more stringent in both
rapidity, transverse momentum or angular separation observables of the
different particles. Here we do not aim at reproducing the setup of
the experimental analyses, but just to provide an indication of how
selection cuts can affect the sensitivity to the dimension-6
effects. We find that our additional cuts lead to a significant
reduction of the statistics and to a weakened sensitivity to the
dimension-6 effects for the angular observables considered
here. Despite the reduction in the sensitivity, the shape difference
in the cos\,$\theta_l^x$ distribution (fig.~
\ref{fig:EFT2_pol+hel_cuts} left) between the dipole and other
operators persists. This shape difference can be measurable as an
asymmetry between positive and negative values of cos\,$\theta_l^x$ as
can be seen in fig.~\ref{fig:EFT2_pol_2bin}.

We also mention here that we examined event samples where the operators were only
allowed to enter in the production of the top quark. Here it was observed that for the $W$ helicity angles, 
eqs.~\ref{eq:Whel} and \ref{eq:Whel3D}, no deviation for the SM was observed. This validates that these angles probe the decay vertex only.


\subsection{CP-violation in single top}
\label{sec:cp-odd}

In this subsection we study possible CP-violating effects in single top production.
In the SM CP violation is too small for baryogenesis, which motivates the search 
for new sources of CP-violation. 
Within the EFT, the coefficient of the $\mathcal{O}_{tW}$
operator can have an imaginary part, leading to a new CP-violating interaction. 
Here we study how large this effect could be
and identify observables sensitive to it.

\begin{figure}[H]
\centering
\begin{subfigure}{0.5\textwidth}
\centering
\includegraphics[scale=0.35,page=7]{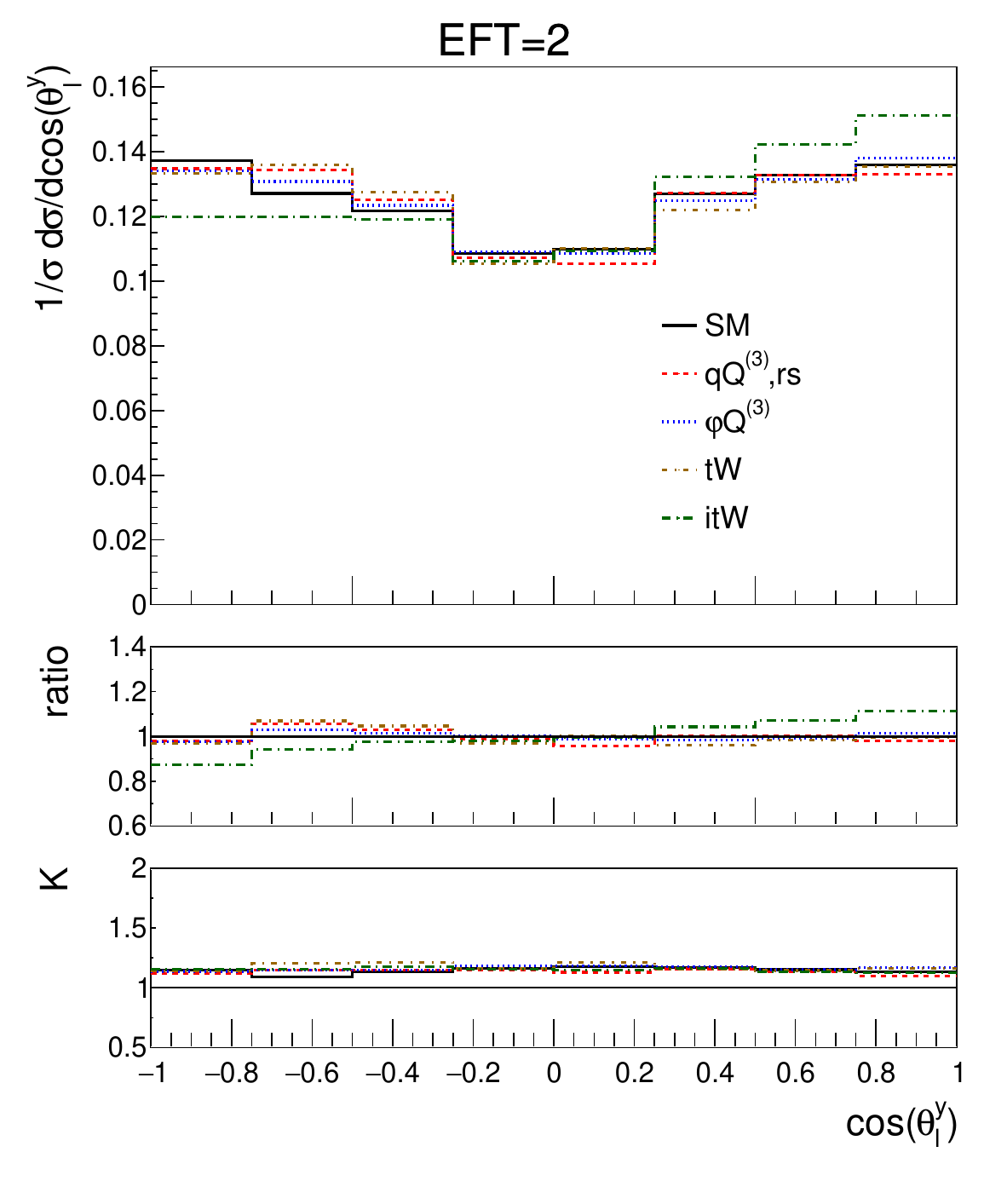}
\end{subfigure}%
\begin{subfigure}{0.5\textwidth}
\centering
\includegraphics[scale=0.35,page=8]{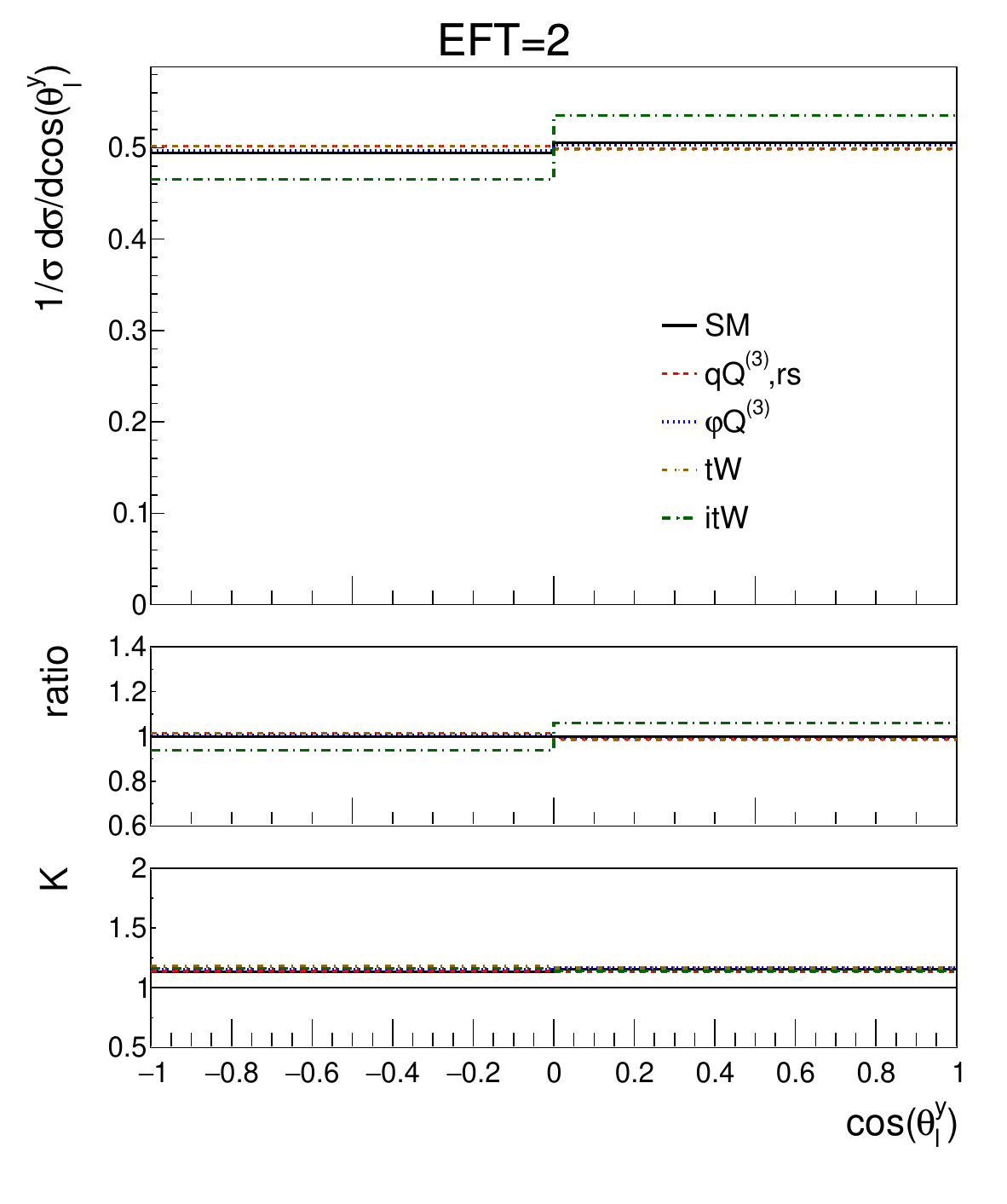}
\end{subfigure}
\caption{The NLO distributions of the top polarisation angle for the SM and the three effective operators of
  interest, together with the imaginary part of $\mathcal{O}_{tW}$ for the couplings values of table \ref{tab:cvalues}. On the left the shape of the distribution can be seen, on
  the right the same distribution is shown in two bins where the
  asymmetry is clearly observed. The ratio shown in the first inset is
  defined as the effect of the operator over the SM, the second inset
  shows the $K$-factor. Here additional cuts are applied on the leptons: $p_T^l > 10$ GeV and $|\eta^l| < 2.47$ and jets: $p_T^j > 20$ GeV and $|\eta^j| < 4.5$.}
\label{fig:poly-complex}
\end{figure}

As discussed in \cite{Aguilar-Saavedra:2014eqa}, the polarisation
angle $\cos{\theta^y_{\ell}}$ defined in eq.~\ref{eq:pol-3D} 
shows a sensitivity to the phase of $\mathcal{O}_{tW}$ coefficient. This can 
indeed be observed in fig.~\ref{fig:poly-complex}, where an asymmetry is
clearly visible, for the imaginary part of the coefficient. The SM, charged current, four-fermion operator and real part of the 
dipole operator show no asymmetry in this distribution.

\begin{figure}[H]
\centering
\begin{subfigure}{0.5\textwidth}
\centering
\includegraphics[scale=0.35,page=7]{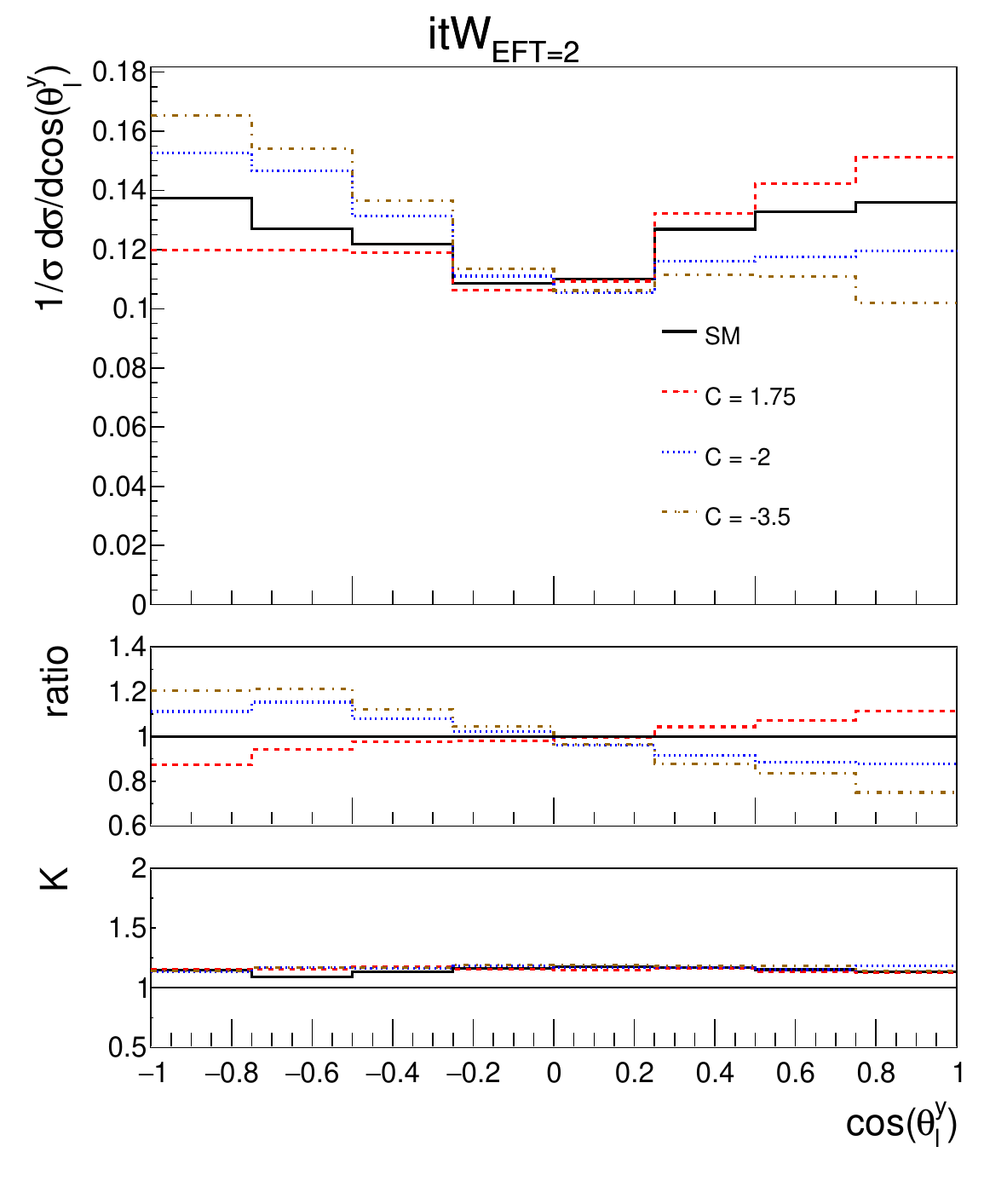}
\end{subfigure}%
\begin{subfigure}{0.5\textwidth}
\centering
\includegraphics[scale=0.35,page=8]{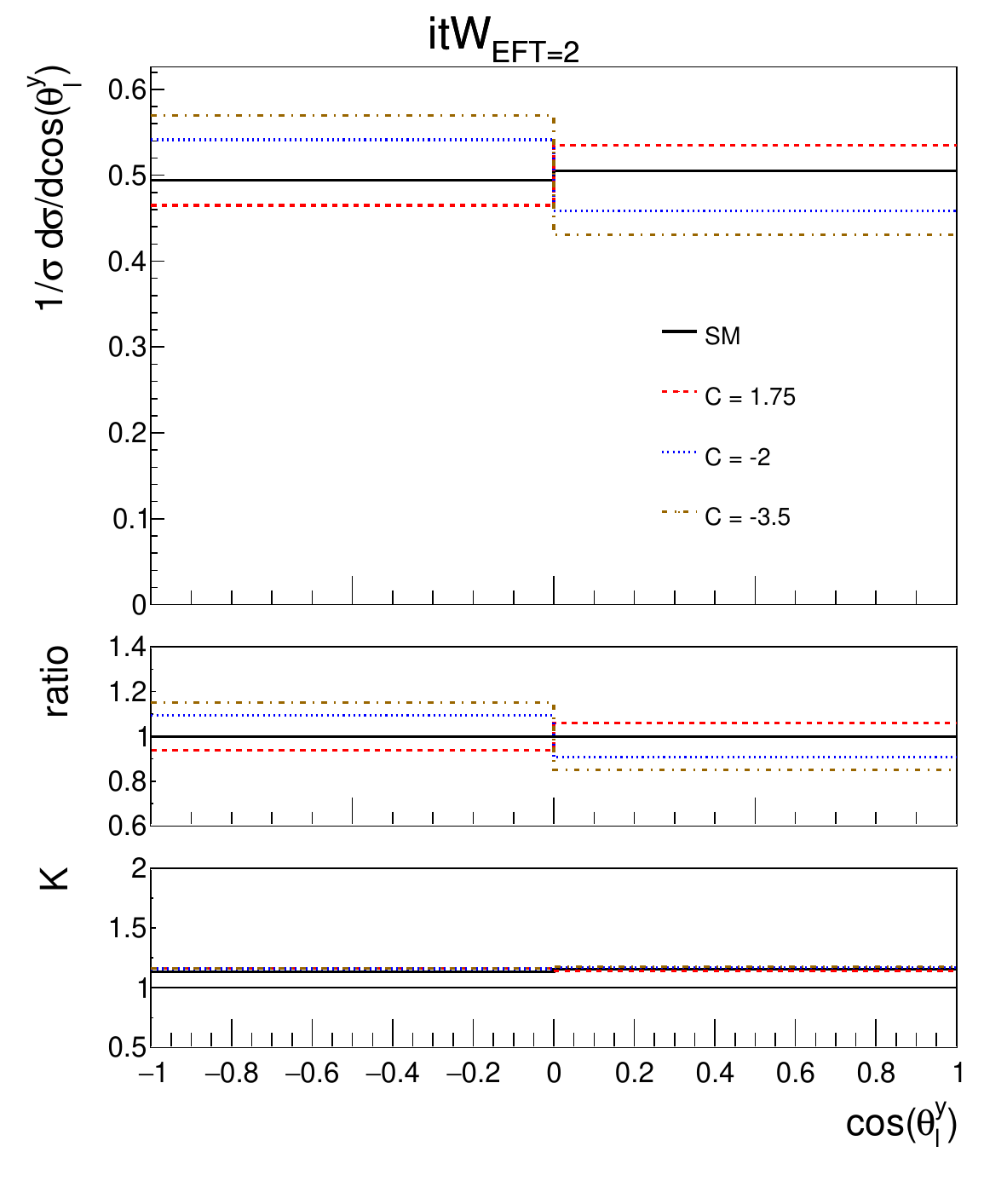}
\end{subfigure}
\caption{The NLO distributions of the top polarisation angle for the SM and different values for the imaginary part of $\mathcal{O}_{tW}$. On the left the shape of the distribution can be seen, on
  the right the same distribution is shown in two bins where the
  asymmetry is clearly observed. The ratio shown in the first inset is
  defined as the effect of the operator over the SM, the second inset
  shows the $K$-factor. Here additional cuts are applied on the leptons: $p_T^l > 10$ GeV and $|\eta^l| < 2.47$ and jets: $p_T^j > 20$ GeV and $|\eta^j| < 4.5$.}
\label{fig:ictW}
\end{figure}

 In order to focus on the effects of the imaginary
part of $C_{tW}$, fig.~\ref{fig:ictW} shows results for a range of coupling values that
are within the current global limits \cite{Cirigliano:2016nyn}. It is
interesting to see that this observable is sensitive to both the size and to the sign
of the coupling for $\mathrm{Im}\,\mathcal{O}_{tW}$. We note here that we 
additionally studied the asymmetry suggested in
\cite{Zhang:2010dr}, but found this to be less sensitive to $\textrm{Im}C_{tW}$ 
than $\cos{\theta^y_{\ell}}$.

\section{Conclusions}
\label{sec:conclude}
Single top production provides an excellent opportunity of probing top quark couplings. The SMEFT is 
a framework which allows us to parametrise deviations from the SM
couplings in a consistent and model-independent way. Predictions in
the SMEFT can be systematically improved by computing higher-order
corrections. In this work we computed for the first time single top
production and decay at NLO in QCD, in the presence of dimension-6
operators. 

We studied the impact of these QCD corrections, both at the inclusive and differential level, and found 
that NLO effects affect both the total rates and the differential distributions in a non-trivial way, 
with different operator contributions receiving different $K$-factors. NLO effects can be large 
and are therefore needed to reliably predict the impact of the dimension-6 operators. We computed all relevant 
contributions at $\mathcal{O}(1/\Lambda^2)$ (and some $\mathcal{O}(1/\Lambda^4)$ terms), and 
examined their relative importance. 

We then included also the decay of the top, examining the validity of the NWA and the impact 
of the top width in computing results for the $Wbj$ final state. We find that the impact of the 
dimension-6 operators on the top width needs to be taken correctly into account to ensure that 
the $Wbj$ and $tj$ cross sections are consistent. We then computed top production and decay at 
NLO matched to the parton shower using the resonance-aware matching within {\sc 
MG5\_aMC}, including off-shell and interference effects. We obtained NLO distributions for both
 the top and its decay products for the SM and a series of benchmarks with non-zero operator
  coefficients. We find that the weak dipole and four-fermion operators can lead to harder tails in the 
  distributions. 
  
In order to fully exploit the power of spin correlations, we explored a series of angular 
observables that can be used to probe new physics couplings in either the production or decay
of the top. These include the so-called polarisation angle and $W$ helicity fractions. We find 
these angular distributions to be sensitive to different operators. The sensitivity becomes weaker  
when we apply cuts on the top decay products, but can still be probed by defining the 
corresponding asymmetries. Finally we considered CP-violating effects coming from the imaginary 
part of the dipole operator coefficient and studied an angular distribution that can be used to 
identify such an interaction.

Our study is an example of using an accurate and realistic simulation framework to 
compute deviations from the SM within SMEFT for a limited number
of operators. Our results can be used in 
combination with the experimental results to obtain reliable constraints on the operator 
coefficients as part of the on-going effort of EFT interpretations of LHC top-quark measurements \cite{AguilarSaavedra:2018nen}.

%

\section*{Acknowledgments}
We would like to thank Fabio Maltoni, Jordy de Vries and Cen Zhang for discussions,
and Rikkert Frederix, Andrew Papanastasiou and Paolo Torrielli for
valuable technical assistance. This work was supported by the 
Netherlands Organisation for Scientific Research
(NWO). EV is supported by a Marie Sk\l{}odowska-Curie Individual
Fellowship of the European Commission's Horizon 2020 Programme under
contract number 704187.



\bibliography{refs.bib}
\end{document}